\documentclass[
 reprint,
 amsmath,amssymb,
]{revtex4-2}

\usepackage{graphicx}
\usepackage{dcolumn}
\usepackage{bm}

\newcommand{\vc}{\mathbf}

\newcommand{\Eq}[1]{Eq.~\eqref{#1}}

\newcommand{\beq}{\begin{equation}}
\newcommand{\eeq}{\end{equation}}

\newcommand{\vect}[1]{\mathbf{#1}}
\newcommand{\E}{\mathrm{e}}
\newcommand{\I}{\mathrm{i}}
\newcommand{\nn}{\nonumber}

\begin{document}

\preprint{APS/123-QED}

\title{Three-body scattering area for particles with infinite or zero scattering length in two dimensions}

\author{Junjie Liang}
\email{junjieliang@pku.edu.cn}
\author{Shina Tan}%
 \email{shinatan@pku.edu.cn}
\affiliation{%
 International Center for Quantum Materials, School of Physics, Peking University, Beijing 100871, China\\
}

\date{April 28, 2024}

\begin{abstract}

We derive the asymptotic expansions of the wave function of three particles having equal mass with finite-range interactions and infinite or zero two-dimensional scattering length colliding at zero energy and zero orbital angular momentum, from which a three-body parameter $D$ is defined.
The dimension of $D$ is length squared, and we call $D$ three-body scattering area.
We find that the ground state energy per particle of a zero-temperature dilute Bose gas with these interactions is approximately $\frac{\hbar^2 D }{6m}\rho^2$, where $\rho$ is the number density of the bosons,  $m$ is the mass of each boson, and $\hbar$ is Planck's constant over $2\pi$. Such a zero-temperature Bose gas is stable at $D> 0$ in the thermodynamic limit, and metastable at $D<0$ in the isotropic harmonic trap if the number of bosons is less than $N_{\mathrm{cr}}\approx 3.6413 \sqrt{\frac{\hbar}{m\omega |D|}}$, where $\omega$ is the angular frequency of the harmonic trap.
If the two-body interaction supports bound states, $D$ typically acquires a negative imaginary part, and we find the relation between this imaginary part and the amplitudes of the pair-boson production processes.
We derive a formula for the three-body recombination rate constant of the many-boson system
in terms of the imaginary part of $D$.
\end{abstract}

\maketitle

\section{Introduction}
The low energy effective interaction of identical bosons or distinguishable particles in two dimensions (2D) is usually
dominated by the 2D scattering length $a$. When the two particles collide with zero total energy
and zero orbital angular momentum, their wave function behaves as
\beq\label{phi}
\phi(\vect \vect s)\propto\ln\frac{s}{a}
\eeq
at $s>r_e$, where $r_e$ is the range of interaction, $\vect s$ is the vector extending from one particle to the other particle,
and $s=|\vect s|$ is the distance between the two particles.
.

In analogy to the scattering length in two-body scatterings, one can define three-body parameters in three-body zero-energy collisions. One of us defined the three-body scattering hypervolume for three identical bosons in three dimensions \cite{tan2008three,zhu2017three}, which was studied by some other authors
\cite{zwerger2019quantum,mestrom2019scattering,mestrom2020van,wang2021three,hu2021first}. The three-body scattering hypervolume has been generalized to three particles with unequal masses \cite{wang2021three}, 
three fermions in three dimensions \cite{wang2021scattering}, two dimensions \cite{wang2022scattering}, and one dimension \cite{wang2023three}. 

Naturally, one would like to study the problem of zero-energy collision of three identical bosons in 2D.
When the 2D scattering length $a$ is finite, the logarithmic behavior of the two-body wave function as shown in \Eq{phi} complicates
the structure of the three-body wave function \cite{kartavtsev2006universal}.
But if $a=\infty$ or $a=0$, the constant $\ln{a}$ dominates the behavior of $\phi(\vect s)$ outside of the range of interaction, and thus
$\phi(\vect s)$ becomes a constant outside of the range of interaction, and the structure of the three-body wave function is simplified
greatly.
The purpose of the present paper is to study such three-body collisions at $a=\infty$ or 0.
In this paper we will define a three-body parameter $D$ for such a system by finding the asymptotic expansions of the three-body wave function
when the three particles are far apart from each other or when one pair of particles is far apart from the remaining particle.
In Ref.~\cite{PhysRevA.100.042707}, this problem was studied assuming a two-channel model;
we will study it assuming more general short-range interactions and we will expand the three-body wave function
to one higher order when the three particles are far apart;
the expansion of the wave function when one pair of particles is far apart from the remaining one
to be derived in this paper seems to be new.
We then study some physical implications of this parameter, including implications for the many-boson systems in 2D.
The parameter $D$ defined and the asymptotic expansions of the three-body wave function derived in this paper
are also applicable to distinguishable particles having equal mass if the three pairwise interaction potentials
are the same.

In another paper we will elucidate the structure of the three-body wave function at zero energy in 2D when $a$ is finite.

In Sec.~\ref{sec:asymptotics}, we study the zero-energy scattering of three identical bosons with short-range interactions in 2D,
fine-tuned such that $a=\infty$ or 0, and derive the asymptotic expansions of the three-body wave function at large interparticle distances, from which the three-body parameter $D$ is defined, which is proportional to the parameter $S_3$ in Ref.~\cite{PhysRevA.100.042707}. The dimension of $D$ is length squared \cite{PhysRevA.100.042707}, in contrast to the three-body scattering hypervolume of identical bosons
in three dimensions whose dimension is length to the fourth power \cite{tan2008three,zhu2017three}.

In Sec.~\ref{sec:energyshift}, we calculate the ground state energy of three identical bosons in a large periodic square, reproducing a result in Ref.~\cite{PhysRevA.100.042707}, and then generalize the result to $N$ bosons. After taking the thermodynamic limit, we find the energy of the Bose gas due to the three-body parameter $D$. The Bose gas
is stable at zero temperature for $D\ge0$ only.
But if $D<0$ the system is unstable in the thermodynamic limit.

In Sec.~\ref{sec:D<0} we study the 2D Bose gas in a harmonic trap at $D<0$.
By numerically solving the Gross-Pitaevskii (GP) equation with $D<0$, we find that the system is metastable if the number of bosons is
less than a critical number. Our result is reminiscent of analogous predictions for the Bose
gases with attractive two-body interactions \cite{ruprecht1995time,adhikari2000numerical}. People have also studied trapped Bose gases with both two-body and three-body couplings \cite{gammal1999trapped,gammal1999improved,gammal2000atomic}, but the harmonically trapped Bose gas with infinite or zero 2D scattering length was not studied to our knowledge.

\section{Asymototic expansions of the three-body wave function\label{sec:asymptotics}}
We consider three identical bosons (or three distinguishable particles with equal mass) with both two-body and three-body interactions. The three-body wave function for zero total momentum, zero total energy, zero orbital angular momentum, and even $y$-parity is
\beq\label{phi3Fourier}
\phi^{(3)}(\vect r_1,\vect r_2,\vect r_3)=\int\frac{d^2k_1d^2k_2}{(2\pi)^4}\phi^{(3)}_{\vect k_1\vect k_2\vect k_3}
\E^{\I(\vect k_1\cdot\vect r_1+\vect k_2\cdot\vect r_2+\vect k_3\cdot\vect r_3)},
\eeq
where $\vect r_1,\vect r_2,\vect r_3$ are the 2D position vectors of the particles,
and $\hbar\vect k_1,\hbar\vect k_2,\hbar\vect k_3$ are their linear momenta satisfying the constraint
\beq
\vect k_1+\vect k_2+\vect k_3\equiv0.
\eeq 
We define a Cartesian coordinate system on the plane on which the particles reside,
such that the position vector $\vect r_i$ is expanded as $\vect r_i=r_{ix}\hat{\vect x}+r_{iy}\hat{\vect y}$, where $\hat{\vect x}$ and $\hat{\vect y}$ are unit vectors long the $+x$ axis and the $+y$ axis respectively, and $r_{ix}$ and $r_{iy}$ are the Cartesian components of $\vect r_i$. We define a symbol $P_y$ such that
\beq\label{Py}
P_y(r_{ix}\hat{\vect x}+r_{iy}\hat{\vect y})\equiv r_{ix}\hat{\vect x}-r_{iy}\hat{\vect y}.
\eeq
When we say that the three-body wave function has even $y$-parity, what we mean is the condition
$$
\phi^{(3)}(P_y\vect r_1,P_y\vect r_2,P_y\vect r_3)=+\phi^{(3)}(\vect r_1,\vect r_2,\vect r_3).
$$
We assume that
\beq\label{phi3approches1}
\phi^{(3)}(\vect r_1,\vect r_2,\vect r_3)\to1
\eeq
when the three pairwise distances between the particles are all large.
By assuming \Eq{phi3approches1}, we can usually uniquely determine the wave function $\phi^{(3)}$ for given interactions,
if there are no two-body bound states.
Those zero-energy three-body wave functions which diverge when the three pairwise distances are large
are usually less important in ultracold atoms physics than the one we study in this paper.
If there is one or more two-body bound states, and we assume that
the resultant bound pair and the third particle fly apart from each other rather than approach each other,
then we can also uniquely determine $\phi^{(3)}$ by also assuming \Eq{phi3approches1}.

We assume that the particles interact with two-body potential $\frac{\hbar^2}{m}U_{\vect k_1\vect k_2\vect k_1'\vect k_2'}$
(where $\vect k_1+\vect k_2\equiv\vect k_1'+\vect k_2'$)
and three-body potential $\frac{\hbar^2}{m}U_{\vect k_1\vect k_2\vect k_3\vect k_1'\vect k_2'\vect k_3'}$ (where
$\vect k_1+\vect k_2+\vect k_3\equiv\vect k_1'+\vect k_2'+\vect k_3'$), such that
the three-body Schr\"{o}dinger equation at zero energy is
\begin{widetext}
\begin{equation}
\begin{split}
    &\frac{q_1^2+q_2^2+q_3^2}{2}\phi^{(3)}_{\vc{q}_1\vc{q}_2\vc{q}_3}+\frac{1}{2}\int_{\vc{k}'}U_{\vc{p}_1\vc{k}'}
    \phi^{(3)}_{\vc{q}_1,-\vc{q}_1/2+\vc{k}',-\vc{q}_1/2+\vc{k}'}+\frac{1}{2}\int_{\vc{k}'}U_{\vc{p}_2\vc{k}'}\phi^{(3)}_{-\vc{q}_2-\vc{k}',\vc{q}_2,-\vc{q}_2/2+\vc{k}'}\\
    &+\frac{1}{2}\int_{\vc{k}'}U_{\vc{p}_3\vc{k}'}\phi^{(3)}_{-\vc{q}_3/2+\vc{k}',-\vc{q}_3/2-\vc{k}',\vc{q}_3} +\frac{1}{6}\int_{\vc{k}_1'\vc{k}_2'}U_{\vc{q}_1\vc{q}_2\vc{q}_3\vc{k}_1'\vc{k}_2'\vc{k}_3'}\phi^{(3)}_{\vc{k}_1'\vc{k}_2'\vc{k}_3'}=0,
\end{split}
\end{equation}
\end{widetext}
where  $\int_{\vc{k}'}\equiv\int\frac{d^2{k'}}{(2\pi)^2}$,
$\int_{\vect k_1'\vect k_2'}\equiv\int\frac{d^2k_1'}{(2\pi)^2}\frac{d^2k_2'}{(2\pi)^2}$,
and
\beq\label{U2body}
U_{\vc{k}\vc{k}'}\equiv U_{\vc{k},-\vc{k},\vc{k}',-\vc{k}'},
\eeq
\begin{subequations}
\begin{align}
&\vc{p}_1\equiv\frac{1}{2}(\vc{q}_2-\vc{q}_3),\\
&\vect p_2\equiv\frac{1}{2}(\vect q_3-\vect q_1),\\
&\vect p_3\equiv\frac{1}{2}(\vect q_1-\vect q_2).
\end{align}
\end{subequations}

We assume the following properties for the interactions.
\begin{itemize}
  \item Spatial translational invariance:  $\vc{k}_1+\vc{k}_2\equiv\vc{k}_1'+\vc{k}_2'$ in $U_{\vc{k}_1\vc{k}_2\vc{k}_{1}'\vc{k}_{2}'}$, and $\vc{k}_1+\vc{k}_2+\vc{k}_3\equiv\vc{k}_1'+\vc{k}_2'+\vc{k}_3'$ in $U_{\vc{k}_1\vc{k}_2\vc{k}_3\vc{k}_{1}'\vc{k}_{2}'\vc{k}_3'}$.
\item Finite range:
the interaction is limited within a finite inter-particle distance $r_e$ in real space, and so
$U_{\vect k_1\vect k_2\vect k_1'\vect k_2'}$ is a smooth function of $\vect k_1,\vect k_2,\vect k_1'$ if we set $\vect k_2'=\vect k_1+\vect k_2-\vect k_1'$.
Similarly, $U_{\vc{k_1}\vc{k_2}\vc{k_3}\vc{k_1'}\vect k_2'\vect k_3'}$ is a smooth function of $\vect k_1,\vect k_2,\vect k_3,\vect k_1',\vect k_2'$ if we set $\vect k_3'=\vect k_1+\vect k_2+\vect k_3-\vect k_1'-\vect k_2'$.
  \item Bosonic symmetry: because of Bose statistics, we can symmetrize $U$ with respect to the incoming (outgoing) momenta without losing generality: $U_{\vc{k}_1\vc{k}_2\vc{k}_{1}'\vc{k}_{2}'}=U_{\vc{k}_2\vc{k}_1\vc{k}_1'\vc{k}_2'}=U_{\vc{k}_1\vc{k}_2\vc{k}_2'\vc{k}_1'}$.
  Similarly, $U_{\vect k_1\vect k_2\vect k_3\vect k_1'\vect k_2'\vect k_3'}$ is symmetric under the interchange of the incoming (outgoing)
  momenta.
\item Hemiticity:  $U_{\vc{k}_1\vc{k}_2\vc{k}_{1}'\vc{k}_{2}'}=U_{\vc{k}_2'\vc{k}_1'\vc{k}_2\vc{k}_1}^*$,
and $U_{\vect k_1\vect k_2\vect k_3\vect k_1'\vect k_2'\vect k_3'}=U_{\vect k_3'\vect k_2'\vect k_1'\vect k_3\vect k_2\vect k_1}^*$.
  \item Rotational invariance: for any rotation $R$ on the 2D plane, $U_{\vc{k}_1\vc{k}_2\vc{k}_1'\vc{k}_2'}=U_{R\vc{k}_1,R\vc{k}_2,R\vc{k}_1',R\vc{k}_2'}$ and $U_{\vc{k}_1\vc{k}_2\vc{k}_3\vc{k}_{1}'\vc{k}_{2}'\vc{k}_3'}=U_{R\vc{k}_1,R\vc{k}_2,R\vc{k}_3,R\vc{k}_1',R\vc{k}_2',R\vc{k}_3'}$.
  \item Galilean invariance: for any 2D vector $\Delta\vect k$,
  $U_{\vc{k}_1+\Delta\vc{k},\vc{k}_2+\Delta\vc{k},\vc{k}_1'+\Delta\vc{k},\vc{k}_2'+\Delta\vc{k}}=U_{\vc{k}_1\vc{k}_2\vc{k}_1'\vc{k}_2'}$ and $U_{\vc{k}_1+\Delta\vc{k},\vc{k}_2+\Delta\vc{k},\vc{k}_3+\Delta\vc{k},\vc{k}_1'+\Delta\vc{k},\vc{k}_2'+\Delta\vc{k},\vc{k}_3'+\Delta\vc{k}}=U_{\vc{k}_1\vc{k}_2\vc{k}_3\vc{k}_{1}'\vc{k}_{2}'\vc{k}_3'}$.
  \item $y$-reversal invariance:
  $U_{\vect k_1\vect k_2\vect k_1'\vect k_2'}=U_{P_y\vect k_1,P_y\vect k_2,P_y\vect k_1',P_y\vect k_2'}$
  and $U_{\vc{k}_1\vc{k}_2\vc{k}_3\vc{k}_{1}'\vc{k}_{2}'\vc{k}_3'}=U_{P_y\vc{k}_1,P_y\vc{k}_2,P_y\vc{k}_3,P_y\vc{k}_1',P_y\vc{k}_2',P_y\vc{k}_3'}$,
  The $y$-reversal invariance is equivalent to the $x$-reversal invariance for rotationally invariant interactions.
\end{itemize}

Note that the leading-order term $1$ in the expansion of $\phi^{(3)}(\vect r_1,\vect r_2,\vect r_3)$
(when the three pairwise distances are all large) has even $y$-parity.
Since the interactions are $y$-reversally invariant, we may focus on the three-body state with even $y$-parity, whose
wave function satisfies
\beq
\phi^{(3)}(P_y\vect r_1,P_y\vect r_2,P_y\vect r_3)=\phi^{(3)}(\vect r_1,\vect r_2,\vect r_3)
\eeq
for all $\vect r_1,\vect r_2,\vect r_3$.
Since the total orbital angular momentum is zero, the three-body wave function also satisfies
\beq
\phi^{(3)}(R\vect r_1,R\vect r_2,R\vect r_3)=\phi^{(3)}(\vect r_1,\vect r_2,\vect r_3)
\eeq
for any 2D rotation $R$. 
So the wave function is an even function under reflection about any axis on the 2D plane.

We define the following Jacobi coordinates:
\begin{subequations}
\begin{equation}
    \vc{s}_1=\vc{r}_2-\vc{r}_3,\quad \vc{s}_2=\vc{r}_3-\vc{r}_1,\quad
    \vc{s}_3=\vc{r}_1-\vc{r}_2,
\end{equation}
\begin{eqnarray}
    \vc{R}_1=\vc{r}_1-\frac{1}{2}(\vc{r}_2+\vc{r}_3),\nonumber\\
    \vc{R}_2=\vc{r}_2-\frac{1}{2}(\vc{r}_3+\vc{r}_1),\nonumber\\
    \vc{R}_3=\vc{r}_3-\frac{1}{2}(\vc{r}_1+\vc{r}_2).
\end{eqnarray}
We also define the hyperradius
\begin{equation}\label{B}
    B=\sqrt{R_i^2+\frac{3}{4}s_i^2}=\sqrt{(s_1^2+s_2^2+s_3^2)/2}.
\end{equation}
\end{subequations}

We will derive the asymptotic expansions of the three-body wave function $\phi^{(3)}$. For this purpose, we will first define two-body special functions.

\subsection{Two-body special functions}

Consider two identical bosons with mass $m$ each, interacting with potential energy $\propto U_{\vect k\vect k'}$ 
in momentum space in the center-of-mass frame, such that when they collide with energy
\beq
E=\frac{\hbar^2k_E^2}{m}
\eeq
in the center-of-mass frame, the wave function $\psi_\vect k$ (where $+\hbar\vect k$ and $-\hbar\vect k$ are the linear momenta of the two bosons in the center-of-mass frame) for their relative motion satisfies
the Schr\"{o}dinger equation
\beq\label{Hpsi2body}
(H\psi)_\vect k=k_E^2\psi_\vect k,
\eeq
where $\hbar^2 H/m$ is the Hamiltonian in the center-of-mass frame, and
\beq\label{HX}
(HX)_\vect k\equiv k^2X_\vect k+\frac12\int\frac{d^2k'}{(2\pi)^2}U_{\vect k\vect k'}X_{\vect k'}
\eeq
for any function $X_\vect k$.
We assume that the interaction is rotationally invariant, $y$-reversally invariant, and bosonic, namely
$U_{\vect k\vect k'}=U_{R\vect k,R\vect k'}=U_{P_y\vect k,P_y\vect k'}
=U_{-\vect k,\vect k'}=U_{\vect k,-\vect k'}$
for any spatial rotation $R$, where $P_y$ is the $y$-reversal defined in \Eq{Py}. 
If the collision energy $E$ is small,
as is often the case for ultracold atoms, we may solve \Eq{Hpsi2body} by introducing two-body special functions
$\phi_\vect k,f_\vect k,g_\vect k,\dots$ that are independent of energy, satisfying the following sequence of equations:
\beq\label{specialfunction1}
(H\phi)_\vect k=0,~~(Hf)_\vect k=\phi_\vect k,~~(Hg)_\vect k=f_\vect k,\dots
\eeq
and expressing the solution to \Eq{Hpsi2body} as a power series:
\beq
\psi_\vect k\propto \phi_\vect k+k_E^2f_{\vect k}+k_E^4g_{\vect k}+\cdots.
\eeq
Equation~\eqref{specialfunction1} does not uniquely fix the definition of the two-body special functions:
even if the collision energy is zero, the scattering wave function is not unique, because the two identical bosons may collide with orbital angular momentum quantum number $l=0,2,4,\cdots$,
and for $l>0$ we also need to specify the $y$-parity of the wave function.
So we may define a sequence of two-body special functions $\phi^{(l,\sigma)}_\vect k,f^{(l,\sigma)}_\vect k,g^{(l,\sigma)}_\vect k,\dots$ for each partial wave channel specified
by the quantum numbers $(l,\sigma)$, where $\sigma=\pm1$ is the $y$-parity:
if $\psi_{P_y\vect k}=\sigma\psi_{\vect k}$ we say that the wave function $\psi_\vect k$ has $y$-parity $\sigma$.
If $l=0$ we must take $\sigma=+1$.
These special functions are defined to satisfy
\begin{subequations}\label{specialfunction2}
\begin{align}
&(H\phi^{(l,\sigma)})_{\vect k}=0,\\
&(Hf^{(l,\sigma)})_{\vect k}=\phi^{(l,\sigma)}_\vect k,\label{specialfunction2f}\\
&(Hg^{(l,\sigma)})_{\vect k}=f^{(l,\sigma)}_\vect k,\\
&\dots\nonumber
\end{align}
and
\begin{align}
&\phi^{(l,+)}_\vect k=\phi^{(l)}_k\cos(l\theta),\\
&\phi^{(l,-)}_\vect k=\phi^{(l)}_k\sin(l\theta),\\
&f^{(l,+)}_\vect k=f^{(l)}_k\cos(l\theta),\\
&f^{(l,-)}_\vect k=f^{(l)}_k\sin(l\theta),\\
&g^{(l,+)}_\vect k=g^{(l)}_k\cos(l\theta),\\
&g^{(l,-)}_\vect k=g^{(l)}_k\sin(l\theta),\\
&\dots\nonumber
\end{align}
\end{subequations}
where $k\equiv|\vect k|$, and $\theta$ is the angle from the $+x$ axis to the direction of $\vect k$,
such that $k_x=k\cos\theta$ and $k_y=k\sin\theta$.
Equations~\eqref{specialfunction2} still do not uniquely determine the special functions.
The overall amplitudes of $\phi^{(l)}_k$ are not yet specified.
Even after fixing the definition of $\phi^{(l,\sigma)}_\vect k$,
we still have some degrees of freedom in the definitions of $f^{(l,\sigma)}_\vect k$, $g^{(l,\sigma)}_\vect k$ etc,
because $\tilde{f}^{(l,\sigma)}_\vect k=f^{(l,\sigma)}_\vect k+c_1\phi^{(l,\sigma)}_\vect k$
and $\tilde{g}^{(l,\sigma)}_\vect k=g^{(l,\sigma)}_\vect k+c_1f^{(l,\sigma)}_\vect k+c_2\phi^{(l,\sigma)}_\vect k$ etc
satisfy
$$
(H\tilde{f}^{(l,\sigma)})_\vect k=\phi^{(l,\sigma)}_\vect k,~~
(H\tilde{g}^{(l,\sigma)})_\vect k=\tilde{f}^{(l,\sigma)}_\vect k,
$$
etc, for any coefficients $c_1$ and $c_2$ etc. These degrees of freedom in the definitions of the special functions
may be called some ``gauge freedom".
The general solution to \Eq{Hpsi2body} is
\beq
\psi_\vect k=\sum_{l\sigma}C_{l,\sigma}(\phi^{(l,\sigma)}_\vect k+k_E^2f^{(l,\sigma)}_\vect k+k_E^4g^{(l,\sigma)}_\vect k+\cdots),
\eeq
where $C_{l,\sigma}$ are any coefficients depending on $(l,\sigma)$.

For later convenience we also define two-body special functions that have well-defined parity for reflection
about an arbitrary unit vector $\hat{\vect n}=\hat{\vect x}\cos\theta_n+\hat{\vect y}\sin\theta_n$:
\begin{subequations}
\begin{align}
\phi^{(l)}_{\hat{\vect n}\vect k}&=\phi^{(l,+)}_\vect k\cos(l\theta_n)+\phi^{(l,-)}_\vect k\sin(l\theta_n)\nonumber\\
&=\phi^{(l)}_k\cos(l\,\mathrm{arg}(\hat{\vect n},\hat{\vect k})),\\
f^{(l)}_{\hat{\vect n}\vect k}&=f^{(l,+)}_\vect k\cos(l\theta_n)+f^{(l,-)}_\vect k\sin(l\theta_n)\nonumber\\
&=f^{(l)}_k\cos(l\,\mathrm{arg}(\hat{\vect n},\hat{\vect k})),\\
g^{(l)}_{\hat{\vect n}\vect k}&=g^{(l,+)}_\vect k\cos(l\theta_n)+g^{(l,-)}_\vect k\sin(l\theta_n)\nonumber\\
&=g^{(l)}_k\cos(l\,\mathrm{arg}(\hat{\vect n},\hat{\vect k})),\\
\dots,\nonumber
\end{align}
\end{subequations}
where $\mathrm{arg}(\hat{\vect n},\hat{\vect k})$ is the angle formed by the unit vectors $\hat{\vect n}$ and $\hat{\vect k}$.
These functions satisfy
\beq
(H\phi^{(l)}_{\hat{\vect n}})_\vect k=0,
~~(Hf^{(l)}_{\hat{\vect n}})_\vect k=\phi^{(l)}_{\hat{\vect n}\vect k},
~~(Hg^{(l)}_{\hat{\vect n}})_\vect k=f^{(l)}_{\hat{\vect n}\vect k},
\dots.\label{Hspecial}
\eeq
We define the Fourier transforms of the above special functions:
\begin{subequations}
\begin{align}
&\phi^{(l)}_{\hat{\vect n}}(\vect s)=\int\frac{d^2k}{(2\pi)^2}\phi^{(l)}_{\hat{\vect n}\vect k}e^{\I\vect k\cdot\vect s},\\
&f^{(l)}_{\hat{\vect n}}(\vect s)=\int\frac{d^2k}{(2\pi)^2}f^{(l)}_{\hat{\vect n}\vect k}e^{\I\vect k\cdot\vect s},\\
&g^{(l)}_{\hat{\vect n}}(\vect s)=\int\frac{d^2k}{(2\pi)^2}g^{(l)}_{\hat{\vect n}\vect k}e^{\I\vect k\cdot\vect s},\\
&\dots.\nonumber
\end{align}
\end{subequations}
For $s$-wave collisions, $l=0$, the special functions do not depend on $\hat{\vect n}$ and we may simply write
$\phi_\vect k\equiv\phi^{(0)}_{\hat{\vect n}\vect k}$,
$f_\vect k\equiv f^{(0)}_{\hat{\vect n}\vect k}$,
$g_\vect k\equiv g^{(0)}_{\hat{\vect n}\vect k}$,
etc,
and
$\phi(\vect s)\equiv\phi^{(0)}_{\hat{\vect n}}(\vect s)$,
$f(\vect s)\equiv f^{(0)}_{\hat{\vect n}}(\vect s)$,
$g(\vect s)\equiv g^{(0)}_{\hat{\vect n}}(\vect s)$,
etc.
For $d$-wave collisions, $l=2$, we write $\phi^{(d)}_{\hat{\vect n}\vect k}\equiv\phi^{(2)}_{\hat{\vect n}\vect k}$,
$f^{(d)}_{\hat{\vect n}\vect k}\equiv f^{(2)}_{\hat{\vect n}\vect k}$, etc,
and $\phi^{(d)}_{\hat{\vect n}}(\vect s)\equiv\phi^{(2)}_{\hat{\vect n}}(\vect s)$,
$f^{(d)}_{\hat{\vect n}}(\vect s)\equiv f^{(2)}_{\hat{\vect n}}(\vect s)$, etc.

Assuming that the interaction is finite-ranged in real space, such that $U_{\vect k\vect k'}$ is a smooth function
of $\vect k$ and $\vect k'$, we may analytically determine the two-body special functions outside of the range
of interaction $r_e$.
If the 2D $s$-wave scattering length $a$ is finite, then $\phi(\vect s)$ satisfies \Eq{phi}.
But in this paper we will be focused on those interactions for which $a=\infty$ or $0$, and so $\phi(\vect s)$
is a constant at $s>r_e$. We choose this constant to be 1 and thereby fix the overall amplitude of $\phi(\vect s)$:
\beq
\phi(\vect s)=1,~~\text{if }s>r_e.\label{phix>re}
\eeq
When $s>r_e$ the special functions $f(\vect s)$ and $g(\vect s)$ etc satisfy
$-\nabla^2 f(\vect s)=\phi(\vect s)$ and $-\nabla^2 g(\vect s)=f(\vect s)$ etc, where $\nabla^2$ is the Laplace operator with respect to $\vect s$.
Solving the first equation we find
\beq
f(\vect s)=-\frac{1}{4}s^2+r_{s*}\ln\frac{s}{x_*},~~\text{if }s>r_e,\label{fx>re}
\eeq
where $r_{s*}$ depends on the details of the interaction, but $x_*$ is an arbitrary length scale.
$r_{s*}$ can be positive, negative, or zero for different interactions.
If $r_{s*}\neq0$, different choices of $x_*$ correspond to different ``gauges" discussed above:
if we add any multiple of $\phi(\vect s)$ to the definition of $f(\vect s)$ we will effectively change the value of $x_*$.
Given the above formula for $f(\vect s)$ we find that
\beq\label{gx>re}
 g(\vc{s})=\frac{1}{64}s^4+\frac{r_{s*}}{4}s^2(1-\ln{\frac{s}{x_*}})+r_{s*}'\ln{\frac{s}{x_*}}
\eeq
if $s>r_e$, where we have fixed the ``gauge freedom" in $g(\vect s)$ by requiring that there is no extra constant
term on the right hand side of \Eq{gx>re}. $r'_{s*}$ is another parameter that depends on the details of the interaction.
$\phi^{(d)}_{\hat{\vect n}}(\vect s)$ satisfies $-\nabla^2\phi^{(d)}_{\hat{\vect n}}(\vect s)=0$ at $s>r_e$.
We can fix the overall amplitude of $\phi^{(d)}_{\hat{\vect n}}(\vect s)$ by requiring that
\beq
\phi^{(d)}_{\hat{\vc{n}}}(\vc{s})=\Big(\frac{s^2}{8}-\frac{4a_d}{\pi s^2}\Big)\cos[2 \mathrm{arg}(\hat{\vc{n}},\hat{\vc{s}})]\label{phidx>re}
\eeq
at $s>r_e$, where $a_d$ is the $d$-wave scattering ``length" having the dimension
of length to the fourth power.

Taking inverse Fourier transformations, we find that at small $\vect k$ we have
\begin{subequations}\label{twobodyspecialsmallk}
\begin{equation}
    \phi_{\vc{k}}=(2\pi)^2\delta(\vc{k})+u_0+u_1\vc{k}^2+O(k^4),\label{phiksmallk}
\end{equation}
\begin{equation}
    f_{\vc{k}}=\pi^2\nabla_{\vc{k}}^2\delta(\vc{k})-2\pi r_{s*}\frac{Z_{k_*}(k)}{k^2}+f_0+O(k^2),\label{fksmallk}
\end{equation}
\begin{equation}
    g_{\vc{k}}=\frac{\pi^2}{16}\nabla_{\vc{k}}^4\delta(\vc{k})-2\pi r_{s*}\frac{Z_{k_*}(k)}{k^4}-2\pi r_{s*}'\frac{Z_{k_*}(k)}{k^2}+g_0+O(k^2)
\end{equation}
\begin{equation}
    \phi^{(d)}_{\hat{\vc{n}}\vc{k}}=-\frac{\pi^2}{2}Q^{(d)}_{\hat{\vc{n}}}(\nabla_{\vc{k}})\delta(\vc{k})+\frac{4a_d}{k^2}Q^{(d)}_{\hat{\vc{n}}}(\vc{k})+O(k^2),\label{phidksmallk}
\end{equation}
\label{TwoBodyWF}
\end{subequations}
where $u_0$, $u_1$, $f_0$, and $g_0$ are parameters that depend on the details of the interaction,
\beq\label{k_*}
k_*=2e^{-\gamma}/x_*,
\eeq
$\gamma=0.5772\cdots$ is Euler's constant,
\begin{equation}
	Q^{(d)}_{\hat{\vc{n}}}(\vc{k})=2(\hat{\vc{n}}\cdot\vc{k})^2-k^2
\end{equation}
is a harmonic polynomial,
\beq
Q^{(d)}_{\hat{\vect n}}(\nabla_\vect k)=2(\hat{\vect n}\cdot\nabla_\vect k)^2-\nabla_\vect k^2,
\eeq
and we have used some generalized functions first introduced in Ref.~\cite{tan2008three} for three dimensions,
but here we define two analogous generalized functions, $\frac{Z_{k_*}(k)}{k^2}$ and
$\frac{Z_{k_*}(k)}{k^4}$, called $Z$ functions, in 2D:
\begin{subequations}
\label{Zfunctions1}
\begin{equation}
    \frac{Z_{k_*}(k)}{k^2}=\frac{1}{k^2}, \text{ if }k>0,
\end{equation}
\begin{equation}
    \int_{k<k_*}\frac{Z_{k_*}(k)}{k^2}\mathrm{d}^2k=0,
\end{equation}
\begin{equation}
    \frac{Z_{k_*}(k)}{k^4}=\frac{1}{k^4},\text{ if }k>0,
\end{equation}
\begin{equation}
    \int_{\text{all }\vect k}\frac{Z_{k_*}(k)}{k^4}\mathrm{d}^2k=0,
\end{equation}
\begin{equation}
    \int_{k<k_*}\frac{Z_{k_*}(k)}{k^4}k^2\mathrm{d}^2k=0.
\end{equation}
\end{subequations}
Just like the Dirac delta function which can be approached by a sequence of ordinary functions,
the $Z$ functions defined here and those defined in Ref.~\cite{tan2008three} can be approached by sequences of ordinary
functions.
For example, the sequence of ordinary functions of 2D vector $\vect k$,
\beq
F_\eta(\vect k)\equiv\left\{\begin{array}{ll}\frac{1}{k^2},&k\ge\eta,\\
-\frac{2}{\eta^2}\ln\frac{k_*}{\eta},&k<\eta,\end{array}\right.
\eeq
approaches $\frac{Z_{k_*}(k)}{k^2}$ when $\eta\to0^+$.
Just like those for the Dirac delta function, there are uncountably infinitely many different sequences of ordinary functions that approach any given $Z$ function.
Some important mathematical properties of these $Z$ functions are shown in Appendix~\ref{sec:Z}.

In general, the small-$\vect k$ expansion of $\phi^{(l)}_{\hat{\vect n}\vect k}$ is of the form
\beq\label{philksmallk}
\phi^{(l)}_{\hat{\vect n}\vect k}\propto Q^{(l)}_{\hat{\vect n}}(\nabla_\vect k)\delta(\vect k)+o(k^{-l-2}),
\eeq
where $Q^{(l)}_{\hat{\vect n}}(\nabla_\vect k)$ is a degree-$l$ harmonic polynomial of $\nabla_\vect k$.
So the leading-order term in this expansion  scales like $k^{-l-2}$.

From Eqs.~\eqref{HX}, \eqref{Hspecial}, and \eqref{twobodyspecialsmallk} we find that at small $\vect k$
\begin{subequations}\label{integralU}
\begin{align}
\frac{1}{2}\int_{\vc{k}'}U_{\vc{k}\vc{k}'}\phi_{\vc{k}'}&=-u_0 k^2-u_1k^4+O(k^6),\label{integralUphi}\\
\frac12\int_{\vect k'}U_{\vect k\vect k'}f_{\vect k'}&=u_0+2\pi r_{s*}+O(k^2),\label{integralUf}\\
\frac12\int_{\vect k'}U_{\vect k\vect k'}\phi^{(d)}_{\hat{\vect n}\vect k'}&=O(k^2).
\end{align}
\end{subequations}

The parameters that appear in the two-body special functions are related to the two-body scattering phase shifts
at small but nonzero collision energies.

For the $s$-wave two-body collision with energy $E=\hbar^2k_E^2/m$ in the center-of-mass frame, where $k_E>0$, 
the wave function is
\beq
\phi_\vect k(E)=\phi_\vect k+k_E^2f_\vect k+k_E^4g_\vect k+\cdots.
\eeq
Fourier transforming the above wave function, we find that
\begin{align}
\phi(\vect s,E)&=\int\frac{d^2k}{(2\pi)^2}\phi_\vect k(E)e^{\I\vect k\cdot\vect s}\nonumber\\
&=\phi(\vect s)+k_E^2f(\vect s)+k_E^4g(\vect s)+\cdots.
\end{align}
At $s>r_e$ we must have
\beq
\phi(\vect s,E)=G\big[J_0(k_E s)\cot\delta_s-Y_0(k_Es)\big],\label{besselswave}
\eeq
where $\delta_s$ is the $s$-wave scattering phase shift, and $G$ is independent of $s$.
Using Eqs.~\eqref{phix>re}, \eqref{fx>re}, and \eqref{gx>re}, we find that
\beq
G=-\frac{\pi}{2}P(k_E)
    \label{normalizationswave}
\eeq
and
\beq\label{deltas}
\cot\delta_s=\frac{2}{\pi}\Big[-\frac{1}{P(k_E)}+\ln\frac{k_E}{k_*}\Big],
\eeq
where
\beq
P(k_E)=r_{s*}k_E^2+r'_{s*}k_E^4+O(k_E^6).
\eeq
The case of $a=\infty$ or 0 studied here corresponds to the case of vanishing parameter $a_{L,d}$ defined in
Ref.~\cite{hammer2010causality} with $L=0$ and $d=2$.
But if the parameter $a_{L,d}$ in Ref.~\cite{hammer2010causality} vanishes,
the usual effective range expansion in Eq.~(15) of Ref.~\cite{hammer2010causality} is no longer applicable,
and that expansion is replaced by our \Eq{deltas} in the case of $s$-wave collisions in 2D.

For the $d$-wave two-body collision with energy $E=\hbar^2k_E^2/m$ in the center-of-mass frame, where $k_E>0$,
the wave function that is even under reflection about the $\hat{\vect n}$ axis (where $\hat{\vect n}$ is any unit vector) is
\beq
\phi^{(d)}_{\hat{\vect n}\vect k}(E)=\phi^{(d)}_{\hat{\vect n}\vect k}+k_E^2f^{(d)}_{\hat{\vect n}\vect k}+\cdots
\eeq
Fourier  transforming  the  above  wave  function,  we  find that
\beq
\phi^{(d)}_{\hat{\vect n}}(\vect s,E)=\phi^{(d)}_{\hat{\vect n}}(\vect s)+k_E^2 f^{(d)}_{\hat{\vect n}}(\vect s)+\cdots.
\eeq
At $s>r_e$ we must have
\begin{align}
\phi^{(d)}_{\hat{\vect n}}(\vect s,E)&=G^{(d)}\big[J_2(k_Es)\cot\delta_d-Y_2(k_Es)\big]\nn\\
&\quad\times\cos[2\mathrm{arg}(\hat{\vect n},\hat{\vect s})],
\end{align}
where $\delta_d$ is the $d$-wave scattering phase shift, and $G^{(d)}$ is independent of $s$.
Using \Eq{phidx>re}, and assuming that $a_d\ne0$, we find that
\beq
G^{(d)}=-k_E^2a_d
\eeq
and
\beq\label{deltad}
k_E^{4}\cot\delta_d=-\frac{1}{a_d}+O(k_E^2).
\eeq
Equation~\eqref{deltad} agrees with Ref.~\cite{hammer2010causality}.

\subsection{Asymptotics of $\phi^{(3)}_{\vc{k_1}\vc{k_2}\vc{k_3}}$ at small momenta}

Hereafter, we let $\vc{q}$'s be small momenta and $q_i$'s scale like $q^1$. The three-body wave functions could be expanded asymptotically in two limits:
\begin{equation}\label{phi111}
    \phi^{(3)}_{\vc{q}_1\vc{q}_2\vc{q}_3}=\sum_{n=-4}^{\infty}T^{(n)}_{\vc{q}_1\vc{q}_2\vc{q}_3},
\end{equation}
\begin{equation}
    \phi_{\vc{k}}^{\vc{q}}\equiv\phi^{(3)}_{\vc{q},-\vc{q}/2+\vc{k},-\vc{q}/2-\vc{k}}=\sum_{n=-2}^{\infty}S_{\vc{k}}^{(n)\vc{q}},\label{phi21}
\end{equation}
where $T^{(n)}_{\vc{q}_1\vc{q}_2\vc{q}_3}$ and $S_{\vc{k}}^{(n)\vc{q}}$ both scale like $q^n$ (including possibly $q^n\ln^j{q}$, $ j=1,2,\cdots$).
Because $\phi^{(3)}_{\vect k_1\vect k_2\vect k_3}$ is rotationally invariant and symmetric under the
interchange of any two momenta, we can show that
\beq
\phi^{-\vect q}_{\vect k}=\phi^{\vect q}_{-\vect k}=\phi^\vect q_\vect k.
\eeq
So
\beq
S^{(n)-\vect q}_{\vect k}=S^{(n)\vect q}_{-\vect k}=S^{(n)\vect q}_\vect k.
\eeq

When $\vect q$ and $\vect k$ are both small the above functions can be further expanded as
\begin{align}
    &T^{(n)}_{\vc{q},-\vc{q}/2+\vc{k},-\vc{q}/2-\vc{k}}=\sum_i t_{\vc{q}\vc{k}}^{(i,n-i)},\label{Texpand}\\
    &S_{\vc{k}}^{(i)\vc{q}}=\sum_j t_{\vc{q}\vc{k}}^{(i,j)},\label{Sexpand}
\end{align}
where $t^{(i,j)}_{\vect q,\vect k}$ scales like $q^ik^j$.

The three-body  Schr\"odinger equation has two special forms:
\begin{equation}
    \frac{q_1^2+q_2^2+q_3^2}{2}\phi^{(3)}_{\vc{q}_1\vc{q}_2\vc{q}_3}=-\Big(\sum_{i=1}^3 \frac{1}{2}\int_{\vc{k}'}U_{\vc{p}_i\vc{k}'}\phi_{\vc{k}'}^{\vc{q}_i}\Big)-U_{\vc{q}_1\vc{q}_2\vc{q}_3}^{\phi},
\label{SE1}
\end{equation}
\begin{equation}
    (H\phi^{\vc{q}})_{\vc{k}}+\frac{3q^2}{4}\phi_{\vc{k}}^{\vc{q}}+W_{\vc{k}}^{\vc{q}}=0,
\label{SE2}
\end{equation}
where
\begin{equation}
    U_{\vc{q}_1\vc{q}_2\vc{q}_3}^{\phi}=\frac{1}{6}\int_{\vc{k}_1'\vc{k}_2'}U_{\vc{q}_1\vc{q}_2\vc{q}_3\vc{k}_1'\vc{k}_2'\vc{k}_3'}\phi^{(3)}_{\vc{k}_1'\vc{k}_2'\vc{k}_3'},
\end{equation}
\begin{eqnarray}
    W_{\vc{k}}^{\vc{q}}&=&\left(\frac{1}{2}\int_{\vc{k}'}U_{-\vc{q}/2+\vc{k},\vc{q}\vc{k}'\vc{k}''}\phi^{(3)}_{-\vc{q}/2-\vc{k},\vc{k}'\vc{k}''}+(\vc{q}\leftrightarrow-\vc{q})\right)\nonumber\\
    &&+U^{\phi}_{-\vc{q}/2+\vc{k},-\vc{q}/2-\vc{k},\vc{q}}.
    \label{Wk}
\end{eqnarray}

Since the three-body potential $U_{\vect q_1,\vect q_2,\vect q_3,\vect k_1',\vect k_2',\vect k_3'}$
depends smoothly on $\vect q_1$ and $\vect q_2$ for $\vect q_1+\vect q_2+\vect q_3\equiv0$, and since this potential and $\phi^{(3)}_{\vect k_1'\vect k_2'\vect k_3'}$
are both rotationally invariant and $y$-reversally invariant,
and since the three-body potential is symmetric with respect to $\vect q_1,\vect q_2,\vect q_3$, we have the following Taylor expansion at small $\vect q_1,\vect q_2,\vect q_3$ if $\vect q_1+\vect q_2+\vect q_3\equiv0$:
\begin{equation}\label{UphiTaylor}
     U_{\vc{q}_1\vc{q}_2\vc{q}_3}^{\phi}=\kappa_0+\kappa_1(q_1^2+q_2^2+q_3^2)+O(q^4).
\end{equation}
Since the three-body potential $U_{-\vect q/2+\vect k,-\vect q/2-\vect k,\vect q,\vect k_1'\vect k_2'\vect k_3'}$ and the wave function $\phi^{(3)}_{\vect k_1'\vect k_2'\vect k_3'}$
are both rotationally invariant and $y$-reversally invariant and the three-body potential is symmetric under the interchange of the outgoing
momenta, we can show that $W^\vect q_\vect k=W^{-\vect q}_{\vect k}=W^{P_y\vect q}_{P_y\vect k}$ and $W^\vect q_\vect k=W^{R\vect q}_{R\vect k}$ for any 2D rotation $R$.
We thus have the following small-$\vect q$ Taylor expansion:
\begin{equation}\label{WqkTaylor}
    W_{\vc{k}}^{\vc{q}}=\sum_{i=0,2,4,\cdots}q^iW_{\hat{\vc{q}}\vc{k}}^{(i)},
\end{equation}
Because \Eq{WqkTaylor} is a Taylor series expansion in nonnegative-integer powers of $q_x$ and $q_y$ (the Cartesian components of $\vect q$), and because $W^{\vect q}_\vect k$ is rotationally invariant and $y$-reversally invariant,
$W^{(i)}_{\hat{\vect q}\vect k}$ can only depend on $\hat{\vect q}$ like
\beq\label{Wsform}
W^{(i)}_{\hat{\vect q}\vect k}=\sum_{l=0,2,4,\cdots,i}c^{(i)}_l(k)\cos(l\arg(\hat{\vect q},\hat{\vect k})),
\eeq
where $c^{(i)}_l(k)$ depends on $i,l, k$ only. In particular, $W^{(0)}_{\hat{\vect q}\vect k}$ does \emph{not} depend on $\hat{\vect q}$ and can be simply written as $W^{(0)}_\vect k\equiv W^{\vect 0}_\vect k$.

Taking the Fourier transformation of $\phi^{(3)}(\vect r_1,\vect r_2,\vect r_3)$ and
using \Eq{phi3approches1},  we find that the leading order of $\phi^{(3)}_{\vc{q}_1\vc{q}_2\vc{q}_3}$ is
 \begin{equation}
     T_{\vc{q}_1\vc{q}_2\vc{q}_3}^{(-4)}=(2\pi)^4\delta(\vc{q}_1)\delta(\vc{q}_2).
     \label{Tm4}
 \end{equation}
 Note that $\vect q_1+\vect q_2+\vect q_3\equiv0$ in $T^{(n)}_{\vect q_1\vect q_2\vect q_3}$.
 Using \Eq{Tm4}, \Eq{SE1} and \Eq{SE2}, we can determine $S^{(-2)}$, $T^{(-2)}$, $S^{(0)}$, $T^{(0)}$, and $S^{(2)}$ step by step.
 
Expanding $T^{(-4)}_{\vect q,-\vect q/2+\vect k,-\vect q/2-\vect k}$ in powers of $q$ [see \Eq{Texpand}], we find
that
\begin{subequations}
\begin{align}
t^{(-2,-2)}_{\vect q\vect k}&=(2\pi)^4\delta(\vect q)\delta(\vect k),\label{t-2-2a}\\
t^{(i,-4-i)}_{\vect q\vect k}&=0,~~i\ne-2.\label{ti,-4-i}
\end{align}
\end{subequations}

\paragraph*{\textbf{Step 1: Determination of $S^{(-2)}.$}} 

The fact that $t^{(-2,-2)}_{\vect q\vect k}$ is nonzero, together with \Eq{Sexpand}, suggests that
$S^{(-2)\vect q}_\vect k$ is nonzero.
Extracting the terms of order $q^{-2}$ in \Eq{SE2}, we get
\begin{equation}\label{HS-2}
    (H S^{(-2)\vc{q}})_{\vc{k}}=0.
\end{equation}
On the other hand, when $\vect k$ is small
\beq\label{S-2boundary}
S^{(-2)\vect q}_\vect k=t^{(-2,-2)}_{\vect q,\vect k}+O(k^0)=(2\pi)^4\delta(\vect q)\delta(\vect k)+O(k^0).
\eeq
The even-$y$-parity solution to \Eq{HS-2} is
\beq
S^{(-2)\vect q}_\vect k=\sum_{l=0,2,4,\dots}\eta^{(l)}_\vect q\phi^{(l)}_{\hat{\vect q}\vect k},
\eeq
where $\eta^{(l)}_\vect q$ does \emph{not} depend on $\vect k$.
Substituting the small-$ k$ expansions of $\phi^{(l)}_{\hat{\vect q}\vect k}$ shown in Eqs.~\eqref{phiksmallk}
and \eqref{phidksmallk} into the above equation, and comparing the result with \Eq{S-2boundary}, we find
that $\eta^{(0)}_\vect q=(2\pi)^2\delta(\vect q)$ and $\eta^{(l)}_\vect q=0$ for $l\ge2$.
So we find 
\beq
    S^{(-2)\vc{q}}_{\vc{k}}=(2\pi)^2\delta(\vc{q})\phi_{\vc{k}}.
\eeq
Expanding $S^{(-2)\vect q}_\vect k$ at small $\vect k$, and using \Eq{phiksmallk}, we find
\begin{subequations}
\begin{equation}
    t_{\vc{q}\vc{k}}^{(-2,-2)}=(2\pi)^4\delta(\vc{q})\delta(\vc{k}),\label{t-2-2b}
\end{equation}
\begin{equation}\label{t-2,0first}
    t_{\vc{q}\vc{k}}^{(-2,0)}=u_0(2\pi)^2\delta(\vc{q}),
\end{equation}
\beq
t^{(-2,2)}_{\vect q\vect k}=u_1k^2(2\pi)^2\delta(\vect q).
\eeq
\end{subequations}
Equation~\eqref{t-2-2b} agrees with \Eq{t-2-2a}, as we expect.

\paragraph*{\textbf{Step 2: Determination of $T^{(-2)}$ and $S^{(0)}$.}} 

Extracting the terms of order $q^0$ in \Eq{SE2}, we get
\begin{equation}
    (HS^{(0)\vc{q}})_{\vc{k}}=-\frac{3q^2}{4}S_{\vc{k}}^{(-2)\vc{q}}-W_{\vc{k}}^{\vect 0}.
\end{equation}
But
\begin{equation}
    -\frac{3q^2}{4}S_{\vc{k}}^{(-2)\vc{q}}=-\frac{3q^2}{4}(2\pi)^2\delta(\vc{q})\phi_{\vc{k}}=0.
\end{equation}
So we have
\beq\label{HS0}
    (HS^{(0)\vc{q}})_{\vc{k}}=-W^{\vect 0}_\vect k.
\eeq
According to \Eq{Wsform}, $W^\vect 0_\vect k$ is rotationally invariant and depends on $k=|\vect k|$ only.
Before symbolically solving \Eq{HS0}, we introduce a function $d_\vect k$ independent of $\vect q$
and satisfying the equation
\begin{equation}
    (Hd)_{\vc{k}}=-W_{\vc{k}}^{\vect 0}.
    \label{dkeq1}
\end{equation}
Since $W^0_\vect k$ is rotationally invariant, we may require $d_\vect k$ to be rotationally invariant, namely $d_\vect k$ depends on $k=|\vect k|$ only.
Even after making such a requirement on $d_\vect k$, the definition of $d_\vect k$ still has a ``gauge freedom":
$\widetilde{d}_\vect k=d_\vect k+\eta\phi_\vect k$ still satisfies \Eq{dkeq1} and is still rotationally invariant
for any constant $\eta$. Under this change of the definition of $d_\vect k$, the small-$\vect k$ expansion of
$d_\vect k$ will be modified: the coefficients of the terms proportional to $\delta(\vect k)$, $1$, $k^2$ etc will be
modified.
To completely fix the definition of $d_\vect k$, we need to expand $d_\vect k$ at small $\vect k$ to the order $k^{-2}$ and specify the coefficient of the term proportional to $\delta(\vect k)$.

Assuming that $d_\vect k$ has been defined, we may solve \Eq{HS0} and obtain
\begin{equation}\label{S0first}
    S_{\vc{k}}^{(0)\vc{q}}=d_{\vc{k}}+A(q)\phi_{\vc{k}}+A^{(d)}(q)\phi^{(d)}_{\hat{\vc{q}}\vc{k}}+\cdots,
\end{equation}
where $A(q)$ and $A^{(d)}(q)$ etc are all of order $q^0$ (including the possibility of some logarithmic dependencies on $q$)
and independent of the direction of $\vect q$. Since the small-$\vect k$ expansion of $\phi^{(d)}_{\hat{\vect q}\vect k}$ contains a term of order $k^{-4}$ [see the first term on the right hand side of \Eq{phidksmallk}],
$A^{(d)}(q)$ must be zero. If $A^{(d)}(q)$ is not zero, the small-$\vect k$ expansion of $S^{(0)\vect q}_\vect k$ would contain
a nonzero term of order $q^0k^{-4}$. But in \Eq{ti,-4-i} we see that $t^{(0,-4)}=0$.
So $A^{(d)}(q)=0$. Similarly the higher partial-wave terms on the right side of \Eq{S0first} are zero.
So we get
\beq\label{S0second}
S^{(0)\vect q}_\vect k=d_{\vect k}+A(q)\phi_{\vect k}.
\eeq

To find the expansion of $d_\vect k$ at small $\vect k$, we need to first study the expansion of $W^0_\vect k$
at small $\vect k$. From \Eq{Wk}, we get
\beq
W^{\vect 0}_\vect k=U^\phi_{\vect k,-\vect k,\vect 0}+\int_{\vect k'}U_{\frac{\vect k}{2},\vect k'}\phi^{(3)}_{-\vect k,\frac{\vect k}{2}+\vect k',\frac{\vect k}{2}-\vect k'}.
\eeq
When $\vect k$ is small, using \Eq{phi21} and \Eq{UphiTaylor} we get
\begin{align}
W^{\vect 0}_\vect k&=(\kappa_0+2\kappa_1k^2)+\int_{\vect k'}U_{\frac{\vect k}{2},\vect k'}\big[S^{(-2)\vect k}_{\vect k'}+S^{(0)\vect k}_{\vect k'}+S^{(2)\vect k}_{\vect k'}\big]\nn\\
&\quad+O(k^4)\nn\\
&=(\kappa_0+2\kappa_1k^2)\nn\\
&\quad+\int_{\vect k'}U_{\frac{\vect k}{2},\vect k'}
\big[(2\pi)^2\delta(\vect k)\phi_{\vect k'}+d_{\vect k'}+A(k)\phi_{\vect k'}\big]\nn\\
&\quad+\int_{\vect k'}U_{\vect 0,\vect k'}S^{(2)\vect k}_{\vect k'}+O(k^4).\label{W0kTaylorfirst}
\end{align}
When $\vect k$ is small, we have Taylor expansion
\begin{equation}\label{Ud}
    \frac{1}{2}\int_{\vc{k}'}U_{\vc{k},\vc{k}'}d_{\vc{k}'}=\chi_0+\chi_1 k^2+O(k^4).
\end{equation}
Substituting the above equation and \Eq{integralUphi} into \Eq{W0kTaylorfirst},
we get
\begin{align}
W^{\vect 0}_\vect k&=(\kappa_0+2\chi_0)+k^2\Big[2\kappa_1+\frac{\chi_1}{2}-\frac{u_0}{2}A(k)\Big]\nn\\
&\quad+\int_{\vect k'}U_{\vect 0,\vect k'}S^{(2)\vect k}_{\vect k'}+O(k^4).
\end{align}
Substituting the above equation and \Eq{HX} into \Eq{dkeq1}, and using \Eq{Ud}, we get
\begin{align}\label{dkeq2}
k^2d_\vect k&=-D+k^2\Big[-2\kappa_1-\frac{3\chi_1}{2}+\frac{u_0}{2}A(k)\Big]\nn\\
&\quad-\int_{\vect k'}U_{\vect 0,\vect k'}S^{(2)\vect k}_{\vect k'}+O(k^4),
\end{align}
where
\begin{equation}
    D\equiv\kappa_0+3\chi_0.
\end{equation}
Solving \Eq{dkeq2} and using the requirement that $d_\vect k$ is rotationally invariant, we get
\begin{align}
d_{\vect k}&=-D\frac{Z_{k_*}(k)}{k^2}+\Big[-2\kappa_1-\frac{3\chi_1}{2}+\frac{u_0}{2}A(k)\Big]\nn\\
&\quad-\frac{1}{k^2}\int_{\vect k'}U_{\vect 0,\vect k'}S^{(2)\vect k}_{\vect k'}+O(k^2),\label{dsmallkfirst}
\end{align}
where we have fixed the ``gauge freedom" in the definition of $d_\vect k$ by specifying the subscript of the $Z$ function $Z_?(k)/k^2$ to be the $k_*$ defined in \Eq{k_*}. Note that according to \Eq{Zk2subscriptchange}, if we change this subscript,
this $Z$ function will change by a coefficient times $\delta(\vect k)$.

Expanding $S^{(0)\vect q}_\vect k$ shown in \Eq{S0second} at small $\vect k$, and using \Eq{phiksmallk} and \Eq{dsmallkfirst}, we get
\begin{subequations}
\begin{equation}
t_{\vc{q}\vc{k}}^{(0,-2)}=-D\frac{Z_{k_*}(k)}{k^2}+A(q)(2\pi)^2\delta(\vc{k}),\label{t0,-2first}
\end{equation}
\beq\label{t00first}
t^{(0,0)}_{\vect q\vect k}=-2\kappa_1-\frac{3\chi_1}{2}+\frac{u_0}{2}A(k)+u_0A(q)
-\frac{1}{k^2}\int_{\vect k'}U_{\vect 0,\vect k'}S^{(2)\vect k}_{\vect k'}.
\eeq
\end{subequations}

Extracting the terms of order $q^0$ on both sides of \Eq{SE1}, we get
\begin{align}
&\frac{q_1^2+q_2^2+q_3^2}{2}T^{(-2)}_{\vect q_1\vect q_2\vect q_3}=-D+u_0\sum_{i=1}^3(2\pi)^2\delta(\vect q_i)p_i^2\nn\\
&=-D+u_0\sum_{i=1}^3(2\pi)^2\delta(\vect q_i)\Big(p_i^2+\frac34q_i^2\Big).
\end{align}
Using the identity $p_i^2+\frac34q_i^2=\frac{q_1^2+q_2^2+q_3^2}{2}$ for $\vect q_1+\vect q_2+\vect q_3\equiv0$,
we get
\begin{equation}
T_{\vc{q}_1\vc{q}_2\vc{q}_3}^{(-2)}=-\frac{2D}{q_1^2+q_2^2+q_3^2}+u_0\sum_{i=1}^3(2\pi)^2\delta(\vc{q}_i).
\end{equation}
Therefore
\begin{align}
T^{(-2)}_{\vc{q},-\vc{q}/2+\vc{k},-\vc{q}/2-\vc{k}}&=-\frac{D}{k^2+\frac{3}{4}q^2}+(2\pi)^2u_0\Big[\delta(\vc{q})\nonumber\\
&\quad+\delta\Big(-\frac{\vc{q}}{2}+\vc{k}\Big)+\delta\Big(-\frac{\vc{q}}{2}-\vc{k}\Big)\Big].\label{T-2}
\end{align}
When $q$ is small we have the $Z$-$\delta$ expansion
\begin{align}\label{Zdeltafirst}
\frac{1}{k^2+3q^2/4}&=\frac{Z_{k_*}(k)}{k^2}-2\pi\Big(\ln\frac{q}{k_{0*}}\Big)\delta(\vect k)
-\frac{3}{4}q^2\frac{Z_{k_*}(k)}{k^4}\nn\\
&\quad+\frac{3\pi}{8}q^2\Big(\ln\frac{q}{k_{0*}}\Big)\nabla_{\vect k}^2\delta(\vect k)+O(q^4\ln^jq)
\end{align}
for some nonnegative integer $j$, where
\begin{equation}\label{k0star}
    k_{0*}\equiv2k_*/\sqrt{3}.
\end{equation}
To understand how to derive the $Z$-$\delta$ expansion, one may refer to an appendix in Ref.~\cite{tan2008three}.
When $q$ is small we also have the Taylor expansion
\begin{align}\label{Zdeltasecond}
&\delta\Big(-\frac{\vc{q}}{2}+\vc{k}\Big)+\delta\Big(-\frac{\vc{q}}{2}-\vc{k}\Big)\nn\\
&=2\delta(\vect k)+\frac{q^2}{4}(\hat{\vect q}\cdot\nabla_\vect k)^2\delta(\vect k)+O(q^4).
\end{align}
Expanding both sides of \Eq{T-2} in powers of $q$ and using \Eq{Zdeltafirst} and \Eq{Zdeltasecond}, we find

\begin{subequations}
\begin{align}
t_{\vc{q}\vc{k}}^{(-2,0)}&=u_0(2\pi)^2\delta(\vc{q}),\label{t-2,0second}\\
t_{\vc{q}\vc{k}}^{(0,-2)}&=-D\frac{Z_{k_*}(k)}{k^2}
    +\Big(2u_0+\frac{D}{2\pi}\ln{\frac{q}{k_{0*}}}\Big)(2\pi)^2\delta(\vc{k}),\label{t0,-2second}\\
t_{\vc{q}\vc{k}}^{(2,-4)}
&=\pi^2 u_0q^2(\hat{\vect q}\cdot\nabla_{\vc{k}})^2\delta(\vc{k})+\frac{3Dq^2}{4}\frac{Z_{k_*}(k)}{k^4}\nn\\
&\quad-\frac{3\pi Dq^2}{8}\Big(\ln\frac{q}{k_{0*}}\Big)\nabla_{\vc{k}}^2\delta(\vc{k}).\label{t2,-4}
\end{align}
\end{subequations}
Equation~\eqref{t-2,0second} agrees with \Eq{t-2,0first}.

Comparing \Eq{t0,-2first} and \Eq{t0,-2second}, we get
\begin{equation}\label{Aq}
 A(q)=2u_0+\frac{D}{2\pi}\ln{\frac{q}{k_{0*}}}.
\end{equation}
Substituting the above formula into \Eq{S0second}, we get
\beq\label{S0third}
S^{(0)\vect q}_\vect k=d_\vect k+\Big(2u_0+\frac{D}{2\pi}\ln{\frac{q}{k_{0*}}}\Big)\phi_\vect k.
\eeq

\paragraph*{\textbf{Steps 3: Determination of $T^{(0)}$ and $S^{(2)}$.}} 
Extracting the terms of order $q^2$ in \Eq{SE2}, we get
\begin{align}\label{HS2}
    (HS^{(2)\vc{q}})_{\vc{k}}&=
    -\frac{3q^2}{4}S_{\vc{k}}^{(0)\vc{q}}-q^2W^{(2)}_{\hat{\vect q}\vc{k}}\nn\\
    &=-\frac{3}{4}q^2A(q)\phi_\vect k+q^2\Big(-\frac34d_\vect k-W^{(2)}_{\hat{\vect q}\vect k}\Big).
\end{align}

Before symbolically  solving \Eq{HS2}, we introduce a function $d^{(2)}_{\hat{\vect q}\vect k}$ independent of $q$ and satisfying the equation
\beq\label{d2eq1}
(Hd^{(2)}_{\hat{\vect q}})_\vect k=-\frac34d_\vect k-W^{(2)}_{\hat{\vect q}\vect k}.
\eeq
Because the right hand side of the above equation contains only $s$-wave and $d$-wave terms
[because of \Eq{Wsform} and the fact that $d_\vect k$ is rotationally invariant]
and is also $y$-reversally invariant,
and because the two-body interaction is rotationally and $y$-reversally invariant,
we may require that $d^{(2)}_{\hat{\vect q}\vect k}$ be of the form
\beq\label{d2eq2}
d^{(2)}_{\hat{\vect q}\vect k}=d^{(2)}_0(k)+d^{(2)}_2(k)\cos(2\arg(\hat{\vect q},\hat{\vect k})),
\eeq
where $d^{(2)}_0(k)$ and $d^{(2)}_2(k)$ depend only on $k=|\vect k|$.
The above requirement still leaves $d^{(2)}_{\hat{\vect q}\vect k}$ with two remaining degrees of ``gauge freedom",
namely 
$$\widetilde{d}^{(2)}_{\hat{\vect q}\vect k}=d^{(2)}_{\hat{\vect q}\vect k}+\eta_s\phi_\vect k+\eta_d\phi^{(d)}_{\hat{\vect q}\vect k}$$
still satisfies \Eq{d2eq1} and \Eq{d2eq2}, for any coefficients $\eta_s$ and $\eta_d$.
To completely fix the definition of $d^{(2)}_{\hat{\vect q}\vect k}$, we will need to expand $d^{(2)}_{\hat{\vect q}\vect k}$ at small $\vect k$ to the order $k^{-2}$ and specify the coefficients of the terms
proportional to $\delta(\vect k)$ and $Q^{(d)}_{\hat{\vect q}}(\nabla_\vect k)\delta(\vect k)$.

Assuming that $d^{(2)}_{\hat{\vect q}\vect k}$ has been defined, we may solve \Eq{HS2} and obtain
\begin{align}\label{S2first}
        S^{(2)\vc{q}}_{\vc{k}}&=
        q^2d^{(2)}_{\hat{\vect q}\vect k}-\frac{3}{4}q^2 A(q)f_{\vc{k}}+\alpha_0(q)\phi_{\vc{k}}+\alpha_2(q)\phi^{(d)}_{\hat{\vc{q}}\vc{k}}\nn\\
        &\quad+\sum_{l=4,6,8,\dots}^\infty\alpha_{l}(q)\phi^{(l)}_{\hat{\vect q}\vect k}.
\end{align}
where $f_\vect k$ were defined in \Eq{Hspecial} and \Eq{fksmallk},
and $\alpha_l(q)$ is of order $q^2$ (including the possibility of some logarithmic dependencies on $q$).
Note that the small-$\vect k$ expansion of $\phi^{(l)}_{\hat{\vect q}\vect k}$ starts with the term that
scales like $k^{-l-2}$ [see \Eq{philksmallk}]. But $t^{(2,-6)}_{\vect q\vect k}=0$ according to \Eq{ti,-4-i}.
Also, $t^{(2,j)}_{\vect q\vect k}=0$ for $j<-6$ because $T^{(n)}_{\vect q_1\vect q_2\vect q_3}=0$ for $n<-4$.
Therefore $\alpha_l(q)=0$ for $l=4,6,8,\dots$, and
\beq\label{S2second}
S^{(2)\vc{q}}_{\vc{k}}=
        q^2d^{(2)}_{\hat{\vect q}\vect k}-\frac{3}{4}q^2 A(q)f_{\vc{k}}+\alpha_0(q)\phi_{\vc{k}}+\alpha_2(q)\phi^{(d)}_{\hat{\vc{q}}\vc{k}}.
\eeq

Substituting \Eq{Aq} and \Eq{S2second} into \Eq{t00first}, and using Eqs.~\eqref{integralU}, we get
\begin{align}
t^{(0,0)}_{\vect q\vect k}=\xi_0+\frac{u_0D}{2\pi}\ln\frac{q}{k_{0*}}
+D\Big(\frac{u_0}{\pi}+\frac{3r_{s*}}{2}\Big)\ln\frac{k}{k_{0*}},
\end{align}
where
\beq
\xi_0\equiv6u_0^2+6\pi u_0r_{s*}-2\kappa_1-\frac{3\chi_1}{2}-\int_{\vect k'}U_{\vect 0\vect k'}d^{(2)}_{\hat{\vect q}\vect k'}.
\eeq
Using the rotational invariance of the two-body interaction and $d^{(2)}_{\hat{\vect q}\vect k}$,
one can show that $\int_{\vect k'}U_{\vect 0\vect k'}d^{(2)}_{\hat{\vect q}\vect k'}$ is independent of $\hat{\vect q}$.

Expanding $S^{(2)\vect q}_\vect k$ at small $\vect k$, and using the fact that $t^{(2,j)}_{\vect q\vect k}=0$
at $j<-4$ [note \Eq{ti,-4-i} and the fact that $T^{(-3)}_{\vect q_1\vect q_2\vect q_3}=0$ and $T^{(n)}_{\vect q_1\vect q_2\vect q_3}=0$ at $n<-4$], we see that the leading-order term in the small-$\vect k$ expansion of $d^{(2)}_{\hat{\vect q}\vect k}$ is of order $k^{-4}$. Writing the leading order term in the small-$\vect k$ expansion of $d^{(2)}_{\hat{\vect q}\vect k}$ as $L_{\hat{\vect q}\vect k}$,
and comparing the leading-order term of the small-$\vect k$ expansion of $S^{(2)\vect q}_\vect k$ with \Eq{t2,-4},
we get
\begin{align}
q^2L_{\hat{\vect q}\vect k}&=\pi^2[\alpha_2(q)+u_0q^2](\hat{\vect q}\cdot\nabla_\vect k)^2\delta(\vect k)\nn\\
&\quad+\Big[\frac{3\pi^2}{2}u_0q^2-\frac{\pi^2}{2}\alpha_2(q)\Big]\nabla_\vect k^2\delta(\vect k)+\frac{3Dq^2}{4}\frac{Z_{k_*}(k)}{k^4}.
\end{align}
The above equation indicates that $\alpha_2(q)$ must be equal to a constant times $q^2$,
because $L_{\hat{\vect q}\vect k}$ is independent of $q=|\vect q|$.
We can exploit the ``gauge freedom" in the definition of $d^{(2)}_{\hat{\vect q}\vect k}$:
if the small-$\vect k$ expansion of $d^{(2)}_{\hat{\vect q}\vect k}$ contains a term
proportional to $(\hat{\vect q}\cdot\nabla_\vect k)^2\delta(\vect k)$, we can add a certain coefficient times
$\phi^{(d)}_{\hat{\vect q}\vect k}$ to $d^{(2)}_{\hat{\vect q}\vect k}$, such that this term is canceled.
According to this particular choice in the definition of $d^{(2)}_{\hat{\vect q}\vect k}$, we must have
\beq
\alpha_2(q)=-u_0q^2,
\eeq
and thus
\beq\label{L}
L_{\hat{\vect q}\vect k}=\frac{3D}{4}\frac{Z_{k_*}(k)}{k^4}+2\pi^2u_0\nabla_\vect k^2\delta(\vect k).
\eeq
Expanding $S^{(2)\vect q}_{\vect k}$ at small $\vect k$ to the next order term, we get

\begin{align}\label{t2,-2first}
t^{(2,-2)}_{\vect q\vect k}&=q^2L'_{\hat{\vect q}\vect k}+\frac{3\pi r_{s*}}{2}q^2A(q)\frac{Z_{k_*}(k)}{k^2}
+\alpha_0(q)(2\pi)^2\delta(\vect k),
\end{align}
where $L'_{\hat{\vect q}\vect k}$ is the next-order term in the small-$\vect k$ expansion of $d^{(2)}_{\hat{\vect q}\vect k}$.

Extracting the terms of order $q^2$ in \Eq{SE1}, and dividing them by $(q_1^2+q_2^2+q_3^2)/2$, we get
\begin{align}
T^{(0)}_{\vect q_1\vect q_2\vect q_3}=\xi_0
&+\frac{3r_{s*}D}{2(q_1^2+q_2^2+q_3^2)}\sum_{i=1}^3q_i^2\ln\frac{q_i}{k_{0*}}\nn\\
&+\sum_{i=1}^3\Big[u_1p_i^2(2\pi)^2\delta(\vect q_i)+\frac{u_0D}{2\pi}\ln\frac{q_i}{k_{0*}}\Big].
\end{align}

At small $\vect q$ we have the following $Z$-$\delta$ expansions:
\begin{subequations}
\begin{align}\label{Zdeltathird}
&\frac{(\vect k-\vect q/2)^2\ln|\vect k-\vect q/2|+(\vect k+\vect q/2)^2\ln|\vect k+\vect q/2|}{k^2+3q^2/4}\nn\\
&=2\ln k+q^2\frac{Z(k)\cos(2\arg(\hat{\vect q},\hat{\vect k}))}{4k^2}
+\frac{q^2}{2}\frac{Z_{k_*}(k)}{k^2}\nn\\
&\quad-q^2\frac{Z_{k_*}(k)\ln k}{k^2}\nn\\
&\quad+\pi q^2\Big[{\lambda}-\ln^2k_*+\ln k_*-\Big(1+\ln\frac43\Big)\ln q+\ln^2q\Big]\nn\\
&\quad\quad\times\delta(\vect k)+O(q^4),
\end{align}
\begin{align}\label{Zdeltafourth}
&\ln|\vect k-\vect q/2|+\ln|\vect k+\vect q/2|\nn\\
&=2\ln k-q^2\frac{Z(k)\cos(2\arg(\hat{\vect q},\hat{\vect k}))}{4k^2}+\frac{\pi q^2}{4}\delta(\vect k)+O(q^4),
\end{align}
\end{subequations}
where
\begin{subequations}
\begin{align}
&\frac{Z(k)\cos(2\arg(\hat{\vect q},\hat{\vect k}))}{k^2}=\frac{\cos(2\arg(\hat{\vect q},\hat{\vect k}))}{k^2},~\text{if }k>0,\\
&\int_{k<\Lambda}\frac{Z(k)\cos(2\arg(\hat{\vect q},\hat{\vect k}))}{k^2}d^2k\equiv0
\end{align}
\end{subequations}
for any $\Lambda>0$,
\begin{subequations}\label{Zfunction2}
\begin{equation}
    \frac{Z_u(k)\ln\frac{k}{v}}{k^2}=\frac{\ln \frac{k}{v}}{k^2},\text{ if }k>0,
\end{equation}
\begin{equation}
    \int_{k<u}\frac{Z_u(k) \ln \frac{k}{v}}{k^2}\mathrm{d}^2k=0,
\end{equation}
\end{subequations}
$u$ and $v$ are arbitrary positive wave numbers,
and
\begin{align}
{\lambda}&=\lim_{T\to+\infty}\frac14+T(1-\ln T)+\frac{\ln^2T}{4}+\int_0^T\frac{t\ln t}{\sqrt{t^2+t+1}}dt\nn\\
&=\frac{1}{12}\Big[9+\pi^2+3\Big(-6+\ln\frac43\Big)\ln\frac43+6\mathrm{Li}_2\Big(-\frac13\Big)\Big]\nn\\
&\approx1.007118.
\end{align}
$\mathrm{Li}_2(z)=\sum_{j=1}^\infty\frac{z^j}{j^2}$ is the polylogarithm function.

Expanding $T^{(0)}_{\vect q,-\vect q/2+\vect k,-\vect q/2-\vect k}$ in powers of $\vect q$, and using Eqs.~\eqref{Zdeltafirst}, \eqref{k0star}, \eqref{Zdeltasecond}, \eqref{Zdeltathird}, and \eqref{Zdeltafourth},
we get
\begin{subequations}
\begin{align}
t_{\vc{q}\vc{k}}^{(-2,2)}&=u_1k^2(2\pi)^2\delta(\vc{q}),
\end{align}\\
\begin{align}
t^{(0,0)}_{\vect q\vect k}&=\xi_0+\frac{u_0D}{2\pi}\ln\frac{q}{k_{0*}}+D\Big(\frac{u_0}{\pi}+\frac{3r_{s*}}{2}\Big)\ln\frac{k}{k_{0*}},
\end{align}\\
\begin{align}
t^{(2,-2)}_{\vect q\vect k}&=-\frac{3r_{s*}Dq^2}{4}\frac{Z_{k_*}(k)(-\frac12+\ln\frac{k}{q})}{k^2}\nn\\
&\quad+\Big(\frac{3r_{s*}D}{16}-\frac{u_0D}{8\pi}\Big)q^2\frac{Z(k)\cos[2\arg(\hat{\vect q},\hat{\vect k})]}{k^2}\nn\\
&\quad+q^2\Big[\frac{3r_{s*}D}{16\pi}\Big({\lambda}-\ln\frac{q}{k_*}-\ln^2\frac{q}{k_*}\Big)+b\Big](2\pi)^2\delta(\vect k),\label{t2,-2second}
\end{align}
\end{subequations}
where
\beq
b\equiv2u_1+\frac{u_0D}{32\pi^2}.
\eeq
Comparing \Eq{t2,-2second} with \Eq{t2,-2first}, and using the ``gauge freedom" in the definition of $d^{(2)}_{\hat{\vect q}\vect k}$, we have
\beq\label{alpha0}
\alpha_0(q)=q^2\Big[b+\frac{3r_{s*}D}{16\pi}\Big({\lambda}-\ln\frac{q}{k_*}-\ln^2\frac{q}{k_*}\Big)\Big]
\eeq
and
\begin{align}\label{L'}
L'_{\hat{\vect q}\vect k}&=\frac{Z_{k_*}(k)(-3\pi r_{s*}u_0-\frac{3r_{s*}D}{4}\ln\frac{k}{k_*'})}{k^2}\nn\\
&\quad+\Big(\frac{3r_{s*}D}{16}-\frac{u_0D}{8\pi}\Big)\frac{Z(k)\cos[2\arg(\hat{\vect q},\hat{\vect k})]}{k^2},
\end{align}
where
\beq
k_*'=2\sqrt{\frac{e}{3}}k_*\approx1.90378k_*.
\eeq
Here we have fixed the ``gauge freedom" in the definition of $d^{(2)}_{\hat{\vect q}\vect k}$ by requiring
that the coefficient of the term containing $\delta(\vect k)$ is zero, when the subscript of the $Z$
function in the first term on the right side of \Eq{L'} is chosen to be $k_*$.

Combining \Eq{L} and \Eq{L'}, we find the small-$\vect k$ expansion of $d^{(2)}_{\hat{\vect q}\vect k}$:
\begin{align}
d^{(2)}_{\hat{\vect q}\vect k}&=\frac{3D}{4}\frac{Z_{k_*}(k)}{k^4}+2\pi^2u_0\nabla_\vect k^2\delta(\vect k)\nn\\
&\quad+\frac{Z_{k_*}(k)(-3\pi r_{s*}u_0-\frac{3r_{s*}D}{4}\ln\frac{k}{k_*'})}{k^2}\nn\\
&\quad+\Big(\frac{3r_{s*}D}{16}-\frac{u_0D}{8\pi}\Big)\frac{Z(k)\cos[2\arg(\hat{\vect q},\hat{\vect k})]}{k^2}\nn\\
&\quad+O(\ln^jk)
\end{align}
for some nonnegative integer $j$.

\paragraph*{\textbf{Summary and Fourier transforms.}} 

The above step-by-step procedure can be continued, but we stop at $S^{(2)}$  and $T^{(0)}$, and sum up all $S^{(n)}$ ($-2\leq n\leq 2$) or all $T^{(n)}$ ($-4\leq n\leq 0$) we have derived, to get the expansions of the three-body wave function in momentum space:

\begin{widetext}

\begin{subequations}
\begin{align}\label{111momentum}
    \phi^{(3)}_{\vc{q}_1\vc{q}_2\vc{q}_3}&=(2\pi)^4\delta(\vc{q}_1)\delta(\vc{q}_2)-\frac{2D}{q_1^2+q_2^2+q_3^2}+\xi_0
+\frac{3r_{s*}D}{2(q_1^2+q_2^2+q_3^2)}\sum_{i=1}^3q_i^2\ln\frac{q_i}{k_{0*}}\nn\\
&\quad+\sum_{i=1}^3\Big[(u_0+u_1p_i^2)(2\pi)^2\delta(\vect q_i)+\frac{u_0D}{2\pi}\ln\frac{q_i}{k_{0*}}\Big]+O(q^2\ln^j q),
\end{align}
where $\vect q_1+\vect q_2+\vect q_3\equiv0$, and
\begin{align}\label{21momentum}
    \phi^{(3)}_{\vc{q},-\vc{q}/2+\vc{k},-\vc{q}/2-\vc{k}}&=\bigg[(2\pi)^2\delta(\vc{q})+2u_0+\frac{D}{2\pi}\ln{\frac{q}{k_{0*}}}+\alpha_0(q)\bigg]\phi_{\vc{k}}-\frac{3}{4}q^2 \Big(2u_0+\frac{D}{2\pi}\ln{\frac{q}{k_{0*}}}\Big)f_{\vc{k}}
    -u_0 q^2\phi^{(d)}_{\hat{\vc{q}}\vc{k}}\nn\\
    &\quad+d_{\vc{k}}+q^2 d_{\hat{\vc{q}}\vc{k}}^{(2)}+O(q^4\ln^{j'} q),
\end{align}
\end{subequations}
for some nonnegative integers $j$ and $j'$, where $\alpha_0(q)$ is shown in \Eq{alpha0}.

Substituting \Eq{111momentum} into \Eq{phi3Fourier}, we find that when the pairwise distances
$|\vect r_1-\vect r_2|$, $|\vect r_2-\vect r_3|$, and $|\vect r_3-\vect r_1|$ go to infinity simultaneously,
the wave function in real space has the ``111 expansion"
\begin{subequations}
\begin{equation}\label{111expansionspace}
    \phi^{(3)}(\vc{r}_1,\vc{r}_2,\vc{r}_3)=1-\frac{D}{4\pi^2 B^2}-\frac{3r_{s*} D}{4\pi^2 B^4}\sum_{i=1}^3 
   \Big[\frac{1}{2}-(\cos 2\theta_i) \ln \cos\theta_i\Big]+O(B^{-6}\ln^jB),
\end{equation}
where $\theta_i=\mathrm{arctan}\big(\frac{2R_i}{\sqrt{3}s_i}\big)$
is the hyperangle, satisfying $\sum_{i=1}^3\cos(2\theta_i)\equiv0$,
and $B$ is the hyperradius defined in \Eq{B}. The first two terms on the right hand side of \Eq{111expansionspace} are consistent with a result in Ref.~\cite{PhysRevA.100.042707}, and the higher order
term ($\sim B^{-4}$) in our \Eq{111expansionspace} seems to be new.

Substituting \Eq{21momentum} into \Eq{phi3Fourier}, we find that when $\vect s$ is fixed, but $R\to\infty$,
the wave function in real space has the ``21 expansion"
\begin{equation}\label{21realspacefirst}
    \phi^{(3)}(\vc{s}/2,-\vc{s}/2,\vc{R})=\Big[1-\frac{D}{4\pi^2 R^2}+\frac{3r_{s*}D}{8\pi^2R^4}\Big(-3+2\ln\frac{R}{x_*}\Big)\Big]\phi(\vect s)-\frac{3D}{4\pi^2R^4}f(\vect s)+O(R^{-6}\ln^{j'}R).
\end{equation}
\end{subequations}

\end{widetext}

If there are two-body bound states, the three incoming free bosons with zero total energy
may recombine into a dimer and a free boson, which fly apart with total kinetic energy equal to the released two-body binding energy. In this case, the 21 expansion in real space is changed to
\begin{widetext}
\begin{equation}\label{21realspacesecond}
    \phi^{(3)}(\vc{s}/2,-\vc{s}/2,\vc{R})=\Phi+\sum_{l=0,2,4,\cdots,l_{\text{max}}}\sum_{\nu=0}^{\nu_l}C_{l\nu}\sum_{i=1}^3\varphi_{l\nu}(\vc{s}_i,\vc{R}_i),
\end{equation}
\end{widetext}
where $\Phi$ is the right hand side of \Eq{21realspacefirst},
$l_{\text{max}}$ is the maximum orbital angular momentum quantum number of the two-body bound state,
and $(\nu_l+1)$ is the number of vibrational states for orbital angular momentum quantum number $l$.
\begin{equation}
    \varphi_{l\nu}(\vc{s},\vc{R})=u_{l\nu}(s)
    H_{l}^{(1)}\bigg(\frac{2}{\sqrt{3}}\kappa_{l\nu} R\bigg)
    \cos\big(l\,\text{arg}(\hat{\vc{s}},\hat{\vc{R}})\big),
\end{equation}
where $H_l^{(1)}(z)$ is the Hankel function of the first kind, whose asymptotic form at large $z$ is \cite{morse1953methods}
\begin{equation}
  H_l^{(1)}(z)= \sqrt{\frac{2}{\pi z}}e^{\mathrm{i}(z-l\pi/2-\pi/4)}\big[1+O(z^{-1})\big],
\end{equation}
and $u_{l\nu}(s)$ is the radial part of the dimer wave function for vibrational quantum number $\nu$ at
orbital angular momentum quantum number $l$, satisfying the two-body Schr\"odinger equation
\begin{align}
    &(-\nabla_\vect s^2+\kappa_{l\nu}^2)u_{l\nu}(s)\cos\big(l\,\text{arg}(\hat{\vc{s}},\hat{\vc{R}})\big)\nn\\
    &+\frac{1}{2}\int d^2s'U(\vc{s},\vc{s}')u_{l\nu}(s')\cos\big(l\,\text{arg}(\hat{\vc{s}}',\hat{\vc{R}})\big)=0.
\end{align}
$U(\vc{s},\vc{s}')$ is the Fourier transform of $U_{\vc{k}\vc{k}'}$ defined in \Eq{U2body}:
\begin{equation}
    U(\vc{s},\vc{s}')=\int \frac{d^2k}{(2\pi)^2}\int\frac{d^2k'}{(2\pi)^2}U_{\vc{k}\vc{k}'}e^{\mathrm{i}\vc{k}\cdot\vc{s}-\mathrm{i}\vc{k}'\cdot\vc{s}'},
\end{equation}
We impose the following normalization condition for $u_{l\nu}(s)$:
\begin{equation}
    \int_0^{\infty} s u_{l\nu}^*(s)u_{l\nu'}(s)ds=\frac{\delta_{\nu\nu'}}{\big(1+\delta_{l,0}\big)\pi\kappa_{l\nu}^2},
\end{equation}
where $\delta_{l,0}$ is the Kronecker delta.
From the conservation of probability, we find that
\begin{equation}\label{ImD}
    \text{Im}{(D)}=-9\sum_{l\nu} \frac{|C_{l\nu}|^2}{\kappa_{l\nu}^2}.
\end{equation}
We stress that Ref.~\cite{PhysRevA.100.042707} already pointed out that the three-body parameter should have an imaginary
part when there are three-body inelastic processes.

The parameter $D$ has important implications for the four-body physics in 2D.
For the four-boson system without two-body interaction and with three-body resonant interaction in 2D, Nishida proved that there is semisuper Efimov effect \cite{nishida2017semisuper}.
The system Nishida studied \cite{nishida2017semisuper} corresponds to the case of a divergent parameter $D$ defined above.

\paragraph*{\textbf{Correction to the wave function if $1/\ln(a/s_0)$ is small but nonzero.}} 
In the above analyses we assumed that the two-body interaction is fine-tuned such that $a=\infty$ or 0.
In real experiments, due to experimental errors of the parameters used to control
the two-body interaction, one can not make $1/\ln(a/s_0)$ to be exactly zero, where $s_0$ is a length
scale which can be chosen to be comparable to $r_e$. One may do perturbative calculations to estimate the
orders of magnitude of the corrections to the two-body zero-energy wave function $\phi(\vect s)$ and the
three-body zero-energy wave function $\phi^{(3)}$ due to a tiny deviation of the two-body interaction
strength from the critical strength at which $a=\infty$ or $0$.
Let the deviation of the two-body interaction strength from such a critical value be $\delta V$, whose precise
definition is not needed in this paper because we only do order-of-magnitude estimates here.
Using first-order perturbation, one can show that
\beq
\ln\frac{a}{s_0}\approx -\frac{j}{\delta V},
\eeq
where $j$ is a positive constant whose value depends on the precise definition of $\delta V$.
In particular, if $\delta V>0$, namely if the two-body interaction is \emph{slightly less attractive} than the critical interaction, $a$ is exponentially small.
If $\delta V<0$, namely if the two-body interaction is \emph{slightly more attractive} than the critical interaction, $a$ is exponentially large.
The two-body zero-energy wave function for $s$-wave collision, $\phi(\vect s)$, 
is changed to $\phi_{\delta V}(\vect s)$ for a nonzero $\delta V$, and
one can do first-order perturbation to show that for a suitable choice of the overall amplitude of $\phi_{\delta V}(\vect s)$ one has
\beq
\phi_{\delta V}(\vect s)-\phi(\vect s)\approx-\frac{\ln(s/s_0)}{\ln(a/s_0)}
\eeq
at $s>r_e$.
Further doing first-order perturbation, one can show that $\phi^{(3)}$ is modified to $\phi^{(3)}_{\delta V}$,
and
\beq
\phi^{(3)}_{\delta V}(\vect r_1,\vect r_2,\vect r_3)-\phi^{(3)}(\vect r_1,\vect r_2,\vect r_3)
\approx-\sum_{i=1}^3\frac{\ln(s_i/s_0)}{\ln(a/s_0)}
\eeq
when the three pairwise distances $s_1,s_2,s_3$ are all much larger than the range of interaction
but still satisfy $|\frac{\ln(s_i/s_0)}{\ln(a/s_0)}|\ll1$.
One can similarly study the correction to the 21 expansion shown in \Eq{21realspacefirst}.
The result is also that the correction is of the order $O[1/\ln(a/s_0)]$.

\section{$N$-body energy and three-body recombination rate\label{sec:energyshift}}
\subsection{Three bosons in the periodic square}
Now we place the three bosons with interactions specified in Sec.~\ref{sec:asymptotics} into a periodic square of side length $L$, and approximately calculate the ground state energy $E_3$ at large $L$, assuming that the two-body interaction does \emph{not} support any bound state.
The three-body ground state wave function $\psi(\vect r_1,\vect r_2,\vect r_3)$ is periodic and is approximately equal to $1$ when the pairwise distances between the bosons are all comparable to $L$. But when $r_e\ll B\ll L$,
\beq\label{psiapprox}
\psi(\vect r_1,\vect r_2,\vect r_3)\approx1-\frac{D}{4\pi^2B^2}
\eeq
unless two bosons are within the range of interaction.
We also have
\beq
-\frac{\hbar^2}{2m}(\nabla_1^2+\nabla_2^2+\nabla_3^2)\psi(\vect r_1,\vect r_2,\vect r_3)=E_3\psi(\vect r_1,\vect r_2,\vect r_3)
\eeq
when the pairwise distances are all larger than $r_e$.
Since $\psi(\vect r_1,\vect r_2,\vect r_3)$ is periodic,
\beq
\int d^2r_1d^2r_2d^2r_3(\nabla_1^2+\nabla_2^2+\nabla_3^2)\psi(\vect r_1,\vect r_2,\vect r_3)=0,
\eeq
where the integral over $\vect r_i$ is carried out over one periodic square.
The above integral may be divided into two domains: $B<B_0$ and $B>B_0$,
where $r_e\ll B_0\ll L$. The integral in the domain $B>B_0$ is
\begin{align}
&\int_{B>B_0} d^2r_1d^2r_2d^2r_3(\nabla_1^2+\nabla_2^2+\nabla_3^2)\psi(\vect r_1,\vect r_2,\vect r_3)\nn\\
&\approx\int_{B>B_0}d^2r_1d^2r_2d^2r_3(-2m E_3/\hbar^2)\psi(\vect r_1,\vect r_2,\vect r_3)\nn\\
&\approx-2mE_3L^6/\hbar^2.
\end{align}
The integral in the domain $B<B_0$ is
$\int_{B<B_0} d^2r_1d^2r_2d^2r_3(\nabla_1^2+\nabla_2^2+\nabla_3^2)\psi(\vect r_1,\vect r_2,\vect r_3)$,
which may be expressed as a surface integral at $B=B_0$, at which $\psi$ satisfies \Eq{psiapprox}.
This surface integral is $2DL^2$. So we have $2DL^2-2mE_3L^6/\hbar^2\approx0$ and
\beq\label{E3a}
E_3\approx\frac{\hbar^2D}{mL^4}.
\eeq
We expect that $E_3$ has the following expansion at large $L$:
\beq\label{E3b}
E_3=\frac{\hbar^2D}{mL^4}+O(L^{-6}\ln^j L)
\eeq
for some integer $j$.
Equation~\eqref{E3a} or \eqref{E3b} was first derived in Ref.~\cite{PhysRevA.100.042707} for a two-channel model.

If the two-body interaction slightly deviates from the critical interaction (at which $a=\infty$ or 0),
one can do first-order perturbation to show that $E_3$ acquires a correction
\beq
\delta E_3=O\Big[-\frac{\hbar^2}{mL^2\ln(a/s_0)}\Big],
\eeq
provided that $|\delta V|$ is small enough such that $|\ln(L/s_0)|\ll|\ln(a/s_0)|$.

\subsection{Thermodynamic limit}
Now we consider $N$ bosons with interactions specified in Sec.~\ref{sec:asymptotics} and take the thermodynamic limit $N\to \infty$, $L\to\infty$, but the number density $\rho=N/L^2$ is fixed.
In the absence of two-body bound states, the ground state energy per particle is
\begin{equation}\label{EN}
    \frac{E}{N}\approx\frac{1}{N}\binom{N}{3}\frac{\hbar^2 D}{mL^4}\approx \frac{\hbar^2 D }{6m}\rho^2
\end{equation}
if the number density $\rho$ is sufficiently small such that the interparticle spacing $\rho^{-1/2}$ is much
larger than both the range of interaction $r_e$ and the length scale $|D|^{1/2}$.

If the two-body interaction slightly deviates from the critical interaction,
one has a small but nonzero value of $|1/\ln(\rho a^2)|$, then in the thermodynamic limit (in which
$|\ln(L/s_0)|\gg|\ln(a/s_0)|$), one may have either \Eq{EN}
or the formula \cite{PhysRevA.3.1067}
\beq\label{EN2}
\frac{E}{N}\approx-\frac{2\pi\hbar^2\rho}{m\ln(\rho a^2)},
\eeq
depending upon whether
$-\frac{2\pi\hbar^2\rho}{m\ln(\rho a^2)}$ is much smaller or much larger than $\frac{\hbar^2 D }{6m}\rho^2$.
But if $-\frac{2\pi\hbar^2\rho}{m\ln(\rho a^2)}$ is comparable to $\frac{\hbar^2 D }{6m}\rho^2$,
we have not verified whether $E/N$ is determined by the results in Ref.~\cite{PhysRevA.82.063610}.

\subsection{Three-body recombination rate}
If the two-body interaction supports two-body bound states, \Eq{EN} may still describe the many-body system
approximately if the system is in a metastable state at zero temperature,
but now $D$ usually has a negative imaginary part shown in \Eq{ImD}.
Within a short time duration $t$, the probability that no recombination occurs is $\exp(-2 |\mathrm{Im}(E)|t/\hbar)\approx 1-2 |\mathrm{Im}(E)|t/\hbar$, and so the probability of one recombination is $2 |\mathrm{Im}(E)|t/\hbar$. Each recombination results in the loss of three low-energy bosons, so
\begin{equation}
    \frac{d\rho}{dt}=-L_3 \rho^3,
\end{equation}
where $\rho$ is the number density of the remaining low-energy bosons, and
\beq
L_3=\frac{\hbar|\mathrm{Im} (D)|}{m}
\eeq
is the three-body recombination rate constant.

\section{2D Bose gas with negative $D$ in a harmonic trap\label{sec:D<0}}
Consider a 2D Bose gas with infinite or zero 2D scattering length in some external potential $V(\vect r)$.
At zero temperature the gas is approximately described by a macroscopic wave function $\Psi(\vect r)$ satisfying
\begin{equation}\label{Psiconstraint}
    \int |\Psi(\vc{r})|^2\mathrm{d}^2r=N.
\end{equation}
$|\Psi(\vect r)|^2$ is the local number density of the bosons.
The energy functional of the gas is approximately
\begin{equation}\label{energyfunctional}
    E=\int \mathrm{d}^2r\left[\frac{\hbar^2}{2m}|\nabla\Psi|^2+V(\vect r)|\Psi|^2+\frac{\hbar^2D}{6m}|\Psi|^6\right].
\end{equation}
The last term on the right hand side of \Eq{energyfunctional} is a natural generalization of \Eq{EN}.
This energy functional is analogous to the one shown in Ref.~\cite{gammal2000atomic}, but we are now considering a system with \emph{attractive} three-body force ($D<0$) and nearly no two-body force ($a=\infty$ or $0$), in 2D.
The system we are studying here is also reminiscent of the one studied in Ref.~\cite{adhikari2000numerical},
in which there is no three-body force but the two-body force may be attractive.

Minimizing the energy functional with respect to $\Psi(r)$ and $\Psi^*(r)$ under the constraint \Eq{Psiconstraint},
we derive a GP equation
\begin{equation}\label{GP}
    -\frac{\hbar^2}{2m}\nabla^2\Psi(\vc{r})+V(\vc{r})\Psi(\vc{r})+\frac{\hbar^2D}{2m}|\Psi(\vc{r})|^4\Psi(\vc{r})=\mu \Psi(\vc{r}),
\end{equation}
where $\mu$ is the chemical potential and also the Lagrange multiplier for imposing the constraint \Eq{Psiconstraint} when minimizing the energy functional. To our knowledge, a GP equation containing the three-body coupling term
was first considered in Refs.~\cite{gammal1999trapped,gammal1999improved,gammal2000atomic}.

Now consider an isotropic harmonic trap
\beq
V(\vect r)=\frac12m\omega^2r^2,
\eeq
where $\omega>0$ is the angular frequency of the trap. If $\Psi(\vect r)$ is also isotropic and real, such that
\beq
\Psi(\vect r)=\Big(\frac{m\omega}{\hbar|D|}\Big)^{1/4}\widetilde{\Psi}(\widetilde{r}),
\eeq
where
\beq
\widetilde{r}=\sqrt{\frac{m\omega}{\hbar}}r,
\eeq
and $\widetilde{\Psi}(\widetilde{r})$ is a dimensionless function,
the GP equation may be rewritten in the dimensionless form
\begin{equation}\label{GPdimensionless}
-\frac{1}{\widetilde{r}}\frac{d}{d\widetilde{r}}\widetilde{r}\frac{d}{d\widetilde{r}}\widetilde{\Psi}(\widetilde{r})+\widetilde{r}^2\widetilde{\Psi}(\widetilde{r})+c \widetilde{\Psi}^5(\widetilde{r})-2\widetilde{\mu}\widetilde{\Psi}(\widetilde{r})=0,
\end{equation}
where
\begin{equation}
\widetilde{\mu}=\frac{\mu}{\hbar\omega}
\end{equation}
and $c=\mathrm{sgn}(D)$. 
If $D>0$, $c=1$.
If $D<0$, $c=-1$.

\begin{center}
\begin{figure}
\includegraphics{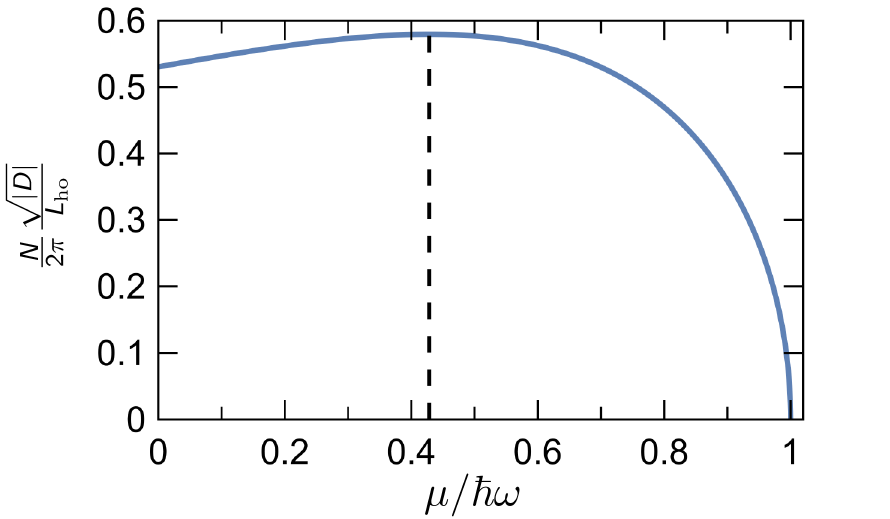}
\caption{\label{critical}The number of bosons $N$ v.s. the chemical potential $\mu$ for a 2D Bose gas with $a=\infty$ or $0$ and
$D<0$ confined in an isotropic harmonic trap with angular frequency $\omega$, at zero temperature. In the graph, $L_\text{ho}\equiv\sqrt{{\hbar}/{m\omega}}$.
The vertical dashed line represents the critical value of $\mu/\hbar\omega$ [see \Eq{mu_cr}]
at which the number $N$ reaches its maximum value [see \Eq{N_cr}].
The segment of the solid-line curve on the left side of the vertical dashed line is not physically realizable
because the 2D gas becomes energetically unstable on that side. We have verified this instability by expanding
the energy functional in \Eq{energyfunctional} around the numerical solution for $\Psi(r)$ [see \Eq{GP} or \Eq{GPdimensionless}], to second order in the
tiny deviations from the solution, and found that the solution is a saddle point but \emph{not} a local minimum
of the energy functional if $\mu<\mu_\text{cr}$.
In Refs.~\cite{akhmediev1999bose,gammal1999improved} it is shown that the $N$-$\mu$ curve for the 3D Bose gas with attractive two-body force and some nonnegative three-body coupling has a similar branch with instability.
We have also verified that if $\hbar\omega>\mu>\mu_\text{cr}$ then the solution
for $\Psi(r)$ [see \Eq{GP} or \Eq{GPdimensionless}] is a local minimum of the energy functional.
}
\end{figure}
\end{center}

If $\widetilde{r}\to\infty$, we have the asymptotic formula \cite{adhikari2000numerical}
\begin{subequations}\label{GPboundary}
\begin{equation}
\widetilde{\Psi}(\widetilde{r})= C \widetilde{r}^{\widetilde{\mu}-1}\big[1+O(\widetilde{r}^{-2})\big]e^{-\widetilde{r}^2/2}.
\end{equation}
If $\widetilde{r}\to0$,
\begin{equation}
\widetilde{\Psi}(\widetilde{r})=\widetilde{\Psi}_0-\frac{(2\widetilde{\mu}\widetilde{\Psi}_0-c\widetilde{\Psi}_0^5)}{4}\widetilde{r}^2+O(\widetilde{r}^4).
\end{equation}
\end{subequations}
For each value of $\widetilde{\mu}$, we numerically solve \Eq{GPdimensionless}, using the conditions in Eqs.~\eqref{GPboundary}, to find the function $\widetilde{\Psi}(\widetilde{r})$, from which we evaluate
the dimensionless number
\begin{equation}
\widetilde{N}\equiv\int_0^{\infty}|\widetilde{\Psi}(\widetilde{r})|^2\widetilde{r}d\widetilde{r}
=\frac{N}{2\pi}\sqrt{\frac{m\omega|D|}{\hbar}}.
\end{equation}
We have done such calculations for $D<0$ only.
Our numerical results are shown in Fig. \ref{critical}. 
For a fixed trap angular frequency $\omega$,
as we gradually increase the dimensionless number $\widetilde{N}$ from zero, the chemical potential decreases gradually from $\hbar\omega$ (the one-body zero-point energy in the 2D harmonic trap),
until $\widetilde{N}$ reaches its maximum value
\beq
\widetilde{N}_\text{cr}=0.579537
\eeq
and $\mu$ reaches its critical value 
\beq\label{mu_cr}
\mu_{\mathrm{cr}}=0.4283\,\hbar\omega.
\eeq
The $\widetilde{N}$-$\widetilde{\mu}$ curve for $0<\widetilde{N}<\widetilde{N}_{\mathrm{cr}}$ is shown in Fig.~\ref{critical};
note that $\mu>\mu_{\mathrm{cr}}$, ie only the segment of the curve to the right of the dashed line in the figure
is physically realizable.
Further increasing $\widetilde{N}$ beyond $\widetilde{N}_{\mathrm{cr}}$ would cause the system to collapse due to a negative $D$.
The maximum number of bosons that can be stably confined in such a trap at zero temperature, for $D<0$, is
\begin{equation}\label{N_cr}
    N_\text{cr}= 2\pi \widetilde{N}_\text{cr}\sqrt{\frac{\hbar}{m\omega|D|}}
    \approx 3.6413\frac{L_\text{ho}}{\sqrt{|D|}},
\end{equation}
where $L_\text{ho}\equiv\sqrt{{\hbar}/{m\omega}}$.
Equation~\eqref{N_cr} is a generalization of those results for the Bose gases with two-body attractive force in 3D \cite{ruprecht1995time} or in 2D \cite{adhikari2000numerical};
it is also reminiscent of those results
for the Bose gases with both two-body attractive force and some \emph{nonnegative} three-body couplings in 3D \cite{akhmediev1999bose,gammal1999improved}.

\section{summary and discussion}
We have derived the asymptotic expansions of the wave function of three identical bosons in 2D with two-body 2D scattering length $a=\infty$ or $0$ , and defined a three-body parameter $D$ whose dimension is length squared. We may call $D$ the \emph{three-body scattering area}.
When there are two-body bound states, $D$ typically acquires a negative imaginary part related to
the probability flux for the production of a bound pair and a free boson.

The three-body scattering area $D$ is a fundamental parameter for the low-energy effective interaction,
and it strongly affects $N$-body physics for $N=3,4,5,\cdots$.

We have evaluated the energy of three such bosons in a large periodic square in terms of $D$ and the side length of the square,
and generalized the result to the dilute 2D Bose gas in the thermodynamic limit.
We then derived a formula for the three-body recombination rate constant in terms of the imaginary part of $D$.

For $D<0$, we have studied the system of $N$ such bosons in an isotropic harmonic trap satisfying the condition
$L_\text{ho}\gg\sqrt{|D|}$, at zero temperature,
using the GP equation, and approximately calculated the maximum value of $N$ for the system to be metastable.

\begin{acknowledgments}
This work was supported by the National Key R$\&$D Program of China (Grants No. 2019YFA0308403 and No. 2021YFA1400902) and the National Natural Science Foundation of China (Grant No. 92365202). We thank Jiansen Zhang for discussions.

\end{acknowledgments}

\appendix
\section{Some mathematical properties of the $Z$ functions defined in this paper\label{sec:Z}}
One can easily show that the $Z$ functions defined in Eqs.~\eqref{Zfunctions1} satisfy the following identities:
\begin{subequations}
\begin{align}
&\frac{Z_{k_1}(k)}{k^2}-\frac{Z_{k_2}(k)}{k^2}=2\pi\Big(\ln\frac{k_2}{k_1}\Big)\delta(\vect k),\label{Zk2subscriptchange}\\
&\frac{Z_{k_1}(k)}{k^4}-\frac{Z_{k_2}(k)}{k^4}=\frac\pi2\Big(\ln\frac{k_2}{k_1}\Big)\nabla_k^2\delta(\vect k),\label{Zk4subscriptchange}
\end{align}
\end{subequations}
for any two positive constants $k_1$ and $k_2$. The Fourier transforms of the $Z$ functions are ordinary functions:
\begin{subequations}
\begin{align}
&\int\frac{d^2k}{(2\pi)^2}\frac{Z_{k_*}(k)}{k^2}e^{\I\vect k\cdot\vect s}=-\frac{1}{2\pi}\ln\frac{s}{x_*},\\
&\int\frac{d^2k}{(2\pi)^2}\frac{Z_{k_*}(k)}{k^4}e^{\I\vect k\cdot\vect s}=\frac{s^2}{8\pi}\Big(-1+\ln\frac{s}{x_*}\Big),
\end{align}
\end{subequations}
where $k_*$ is related to $x_*$ in \Eq{k_*}. One can also verify the inverse Fourier transformations:
\begin{subequations}
\begin{align}
&\int\Big(-\frac1{2\pi}\ln\frac{s}{x_*}\Big)e^{-\I\vect k\cdot\vect s}d^2s=\frac{Z_{k_*}(k)}{k^2},\\
&\int\frac{s^2}{8\pi}\Big(-1+\ln\frac{s}{x_*}\Big)e^{-\I\vect k\cdot\vect s}d^2s=\frac{Z_{k_*}(k)}{k^4},
\end{align}
\end{subequations}

One can also easily show that the $Z$ function defined in Eqs.~\eqref{Zfunction2} satisfies
\beq
\frac{Z_{k_1}(k)\ln\frac{k}{v}}{k^2}-\frac{Z_{k_2}(k)\ln\frac{k}{v}}{k^2}=\pi\Big(\ln\frac{k_2}{k_1}\Big)\Big(\ln\frac{k_1k_2}{v^2}\Big)\delta(\vect k),
\eeq
for any three positive constants $k_1$, $k_2$, and $v$.
We can also calculate the Fourier transform of the $Z$ function defined in Eqs.~\eqref{Zfunction2}:
\beq
\int\frac{d^2k}{(2\pi)^2}\frac{Z_u(k)\ln\frac{k}{v}}{k^2}e^{\I\vect k\cdot\vect s}=
\frac{1}{4\pi}\big[\ln^2(e^\gamma vs/2)-\ln^2(v/u)\big].
\eeq
One can verify the inverse transformation:
\beq
\int\frac{1}{4\pi}\big[\ln^2(e^\gamma vs/2)-\ln^2(v/u)\big]e^{-\I\vect k\cdot\vect s}d^2s=\frac{Z_u(k)\ln\frac{k}{v}}{k^2}.
\eeq

\nocite{*}

\bibliography{apssamp}

\end{document}